\documentstyle[12pt]{article}

\catcode`\@=11
\def\marginnote#1{}
\newcount\hour
\newcount\minute
\newtoks\amorpm
\hour=\time\divide\hour by60
\minute=\time{\multiply\hour by60 \global\advance\minute
by-\hour}\edef\standardtime{{\ifnum\hour<12
\global\amorpm={am}%
        \else\global\amorpm={pm}\advance\hour by-12 \fi
        \ifnum\hour=0 \hour=12 \fi
        \number\hour:\ifnum\minute<10
0\fi\number\minute\the\amorpm}}
\edef\militarytime{\number\hour:\ifnum\minute<10
0\fi\number\minute}

\def\draftlabel#1{{\@bsphack\if@filesw {\let\thepage\relax
   \xdef\@gtempa{\write\@auxout{\string
      \newlabel{#1}{{\@currentlabel}{\thepage}}}}}\@gtempa
   \if@nobreak \ifvmode\nobreak\fi\fi\fi\@esphack}
        \gdef\@eqnlabel{#1}}
\def\@eqnlabel{}
\def\@vacuum{}
\def\draftmarginnote#1{\marginpar{\raggedright\scriptsize\tt#1}}
\def\draft{\oddsidemargin -.5truein
        \def\@oddfoot{\sl preliminary draft \hfil
        \rm\thepage\hfil\sl\today\quad\militarytime}
        \let\@evenfoot\@oddfoot \overfullrule 3pt
        \let\label=\draftlabel
        \let\marginnote=\draftmarginnote

\def\@eqnnum{(\theequation)\rlap{\kern\marginparsep\tt\@eqnlabel}%
\global\let\@eqnlabel\@vacuum}  }

\def\numberbysection{\@addtoreset{equation}{section}
        \def\theequation{\thesection.\arabic{equation}}}

\def\underline#1{\relax\ifmmode\@@underline#1\else
 $\@@underline{\hbox{#1}}$\relax\fi}

\catcode`@=12
\relax

\numberbysection

\advance \topmargin by -\headheight
\advance \topmargin by -\headsep

\evensidemargin \oddsidemargin
\def\lab#1{\label{#1}}

\def\br{\begin{eqnarray}}
\def\er{\end{eqnarray}}
\def\be{\begin{equation}}
\def\ee{\end{equation}}

\def\({\left(}
\def\){\right)}

\newcommand{\bi}[1]{\bibitem{#1}}

\relax

\def\a{\alpha}

\def\b{\beta}

\def\d{\delta}
\def\D{\Delta}

\def\g{\gamma}

\def\l{\lambda}
\def\L{\Lambda}

\def\pa{\partial}

\def\ra{\rightarrow}
\def\s{\sigma}

\def\tp0{\Theta_{+}^{(0)}}
\def\tm0{\Theta_{-}^{(0)}}

\def\vp{\varphi}

\def\f#1#2#3 {f^{#1#2}_{#3}}

\def\win1{{\sf w_{1+\infty}}}

\def\Win1{{\sf W_{1+\infty}}}

\def\rlx{\relax\leavevmode}
\def\inbar{\vrule height1.5ex width.4pt depth0pt}
\def\IZ{\rlx\hbox{\sf Z\kern-.4em Z}}
\def\IR{\rlx\hbox{\rm I\kern-.18em R}}
\def\IC{\rlx\hbox{\,$\inbar\kern-.3em{\rm C}$}}
\def\IN{\rlx\hbox{\rm I\kern-.18em N}}
\def\IO{\rlx\hbox{\,$\inbar\kern-.3em{\rm O}$}}
\def\IP{\rlx\hbox{\rm I\kern-.18em P}}
\def\IQ{\rlx\hbox{\,$\inbar\kern-.3em{\rm Q}$}}
\def\IF{\rlx\hbox{\rm I\kern-.18em F}}
\def\IG{\rlx\hbox{\,$\inbar\kern-.3em{\rm G}$}}
\def\IH{\rlx\hbox{\rm I\kern-.18em H}}
\def\II{\rlx\hbox{\rm I\kern-.18em I}}
\def\IK{\rlx\hbox{\rm I\kern-.18em K}}
\def\IL{\rlx\hbox{\rm I\kern-.18em L}}
\def\one{\hbox{{1}\kern-.25em\hbox{l}}}
\def\0#1{\relax\ifmmode\mathaccent"7017{#1}%
B        \else\accent23#1\relax\fi}

\def\EPC#1#2#3{{\sl Eur. Phys. J.} {\bf C#1} (#2) #3}
                \def\JHEP#1#2#3{{\sl JHEP} {\bf#1} (#2) #3}
                \def\PRL#1#2#3{{\sl Phys. Rev. Lett.} {\bf#1} (#2) #3}
                \def\NPB#1#2#3{{\sl Nucl. Phys.} {\bf B#1} (#2) #3}
                
                \def\CMP#1#2#3{{\sl Commun. Math. Phys.} {\bf #1} (#2) #3}
                \def\PRD#1#2#3{{\sl Phys. Rev.} {\bf D#1} (#2) #3}

                \def\PLB#1#2#3{{\sl Phys. Lett.} {\bf #1B} (#2) #3}
                \def\JMP#1#2#3{{\sl J. Math. Phys.} {\bf #1} (#2) #3}

                \def\AoP#1#2#3{{\sl Annals Phys.} {\bf #1} (#2) #3}
                
                \def\RMP#1#2#3{{\sl Rev. Mod. Phys.} {\bf #1} (#2) #3}
                \def\PR#1#2#3{{\sl Phys. Reports} {\bf #1} (#2) #3}

                \def\IJMPA#1#2#3{{\sl Int. J. Mod. Phys.} {\bf A#1} (#2) #3}

                \def\JPA#1#2#3{{\sl J. Physics} {\bf A#1} (#2) #3}

                \def\JPIV#1#2#3{{\sl J. Phys. IV} {\bf #1} (#2) #3}
                \def\ibid#1#2#3{{\sl ibid} {\bf #1} (#2) #3}

                \def\a{\alpha}
                \def\b{\beta}

                \def\d{\delta}
                \def\D{\Delta}
                
                \def\g{\gamma}
                
                \def\vp{\varphi}

                \def\/{\frac}

                \def\l{\lambda}
                \def\L{\Lambda}

                \def\pa{\partial}

                \def\ra{\rightarrow}
                
                \def\vp{\varphi}
                
                \def\s{\sigma}

                \def\({\Big(}
                \def\){\Big)}
                \def\[{\Big[}
                \def\]{\Big]}

                \def\rlx{\relax\leavevmode}
                \def\inbar{\vrule height1.5ex width.4pt depth0pt}
                \def\IZ{\rlx\hbox{\sf Z\kern-.4em Z}}
                \def\IR{\rlx\hbox{\rm I\kern-.18em R}}
                \def\IC{\rlx\hbox{\,$\inbar\kern-.3em{\rm C}$}}
                \def\IN{\rlx\hbox{\rm I\kern-.18em N}}
                \def\IO{\rlx\hbox{\,$\inbar\kern-.3em{\rm O}$}}
                \def\IP{\rlx\hbox{\rm I\kern-.18em P}}
                \def\IQ{\rlx\hbox{\,$\inbar\kern-.3em{\rm Q}$}}
                \def\IF{\rlx\hbox{\rm I\kern-.18em F}}
                \def\IG{\rlx\hbox{\,$\inbar\kern-.3em{\rm G}$}}
                \def\IH{\rlx\hbox{\rm I\kern-.18em H}}
                \def\II{\rlx\hbox{\rm I\kern-.18em I}}
                \def\IK{\rlx\hbox{\rm I\kern-.18em K}}
                \def\IL{\rlx\hbox{\rm I\kern-.18em L}}
                \def\one{\hbox{{1}\kern-.25em\hbox{l}}}
                \def\0#1{\relax\ifmmode\mathaccent"7017{#1}%
                B        \else\accent23#1\relax\fi}
                
\begin{document}
                \begin{titlepage}
                \begin{center}
                  {\large\bf Bosonization and generalized Mandelstam soliton operators}
                \end{center}

                \begin{center}
                Harold Blas
                \par \vskip .1in
Departamento de Matem\'atica - ICET\\
Universidade Federal de Mato Grosso (UFMT)\\
 Av. Fernando Correa, s/n, Coxip\'o \\
78060-900, Cuiab\'a - MT - Brazil
\end{center}

\begin{abstract}
\vspace{.2 cm}

The generalized massive Thirring model (GMT) with three fermion
species is bosonized in the context of the functional integral and
operator formulations and shown to be equivalent to a generalized
sine-Gordon model (GSG) with three interacting soliton species.
The generalized Mandelstam soliton operators are constructed and
the fermion-boson mapping is established through a set of
generalized bosonization rules in a quotient positive definite
Hilbert space of states. Each fermion species is mapped to its
corresponding soliton in the spirit of particle/soliton duality of
Abelian bosonization. In the semi-classical limit one recovers the
so-called $SU(3)$ affine Toda model coupled to matter fields (ATM)
from which the classical GSG and GMT models were recently derived
in the literature.  The intermediate ATM like effective action
possesses some spinors resembling the higher grading fields of the
ATM theory which have non-zero chirality. These  fields are shown
to disappear from the physical spectrum, thus  providing a bag
model like confinement mechanism and leading to the appearance of
the massive fermions (solitons). The ordinary MT/SG duality turns
out to be related to each $SU(2)$ sub-group. The higher rank Lie
algebra extension is also discussed.
\end{abstract}




                \end{titlepage}

                \section{Introduction}

                A remarkable property which was exploited in the study of
                two-dimensional field theories is related to
                the possibility of transforming Fermi fields into Bose fields, and
                vice versa (see e.g. \cite{stone} and references therein). The existence of
                such a transformation, called  {\sl bosonization}, provided in
                the last years a powerful tool to   obtain nonperturbative information
                in two-dimensional field theories \cite{abdalla}.

               In this context, an  important question is related to the
                multi-flavor extension of the well known massive Thirring (MT)
                and sine-Gordon relationship (SG)\cite{coleman}. In \cite{jmp, jhep} it has been shown through the ``symplectic
                quantization'' and the so-called master Lagrangian approaches that the   generalized  massive Thirring model (GMT) is equivalent to the
                generalized sine-Gordon model (GSG) at the classical level; in particular, the mappings between spinor bilinears of the GMT theory and exponentials of the GSG fields were
                established on shell and the various soliton/particle correspondences were
                uncovered.

                The path-integral version of Coleman's proof of the equivalence between the MT and  SG
                models has been derived in \cite{naon}. In the intermediate process a Lagrangian of the
                so-called $su(2)$ affine Toda model coupled to matter (ATM) \cite{jhep}
                plus a free scalar  appears as a total effective Lagrangian which provides an equivalent
                generating functional to the massive Thirring model after suitable field  redefinitions.
                We generalize the aforementioned result to establish a
                relationship between the  $N_{f}[=3=$ number of positive roots
                of $su(3)$] fermion  GMT and $N_{f}$ boson GSG
                models. Actually, the $U(1)$ GMT currents satisfy a constraint
                and the SG type fields satisfy a
                linear relationship. It is shown that in the $SU(3)$ construction, by taking a
                convenient limiting procedure, each $SU(2)$ sub-group corresponds to
                the ordinary MT/SG duality.

                Earlier attempts used
                nonlinear nonlocal realizations of non-Abelian
                symmetries resorting to $N$ scalar fields \cite{halpern, banks}, in this
                way extending the massive Abelian bosonization
                \cite{coleman}. In this approach the global
                non-Abelian symmetry of the fermions
                is not manifest and the off-diagonal bosonic currents
                become non-local. In Witten's non-Abelian bosonization these
                difficulties were overcome providing manifest global symmetry in the
                 bosonic sector \cite{witten1}. In these
                 developments the appearance of solitons in the
                 bosonized model, which generalizes the
                 sine-Gordon solitons, to our knowledge has not been fully
                 explored; however in Ref. \cite{park} the free massive
                 fermions are considered. The interacting
                 multi-flavor massive fermions deserves a
                 consideration in the spirit of the particle/soliton duality of the
                 Abelian bosonization.

                We perform the bosonization of the GMT model following a hybrid of the
                operator and  functional formalisms in which some auxiliary fields are
                introduced in order to recast the Lagrangian in quadratic form in the
                Fermi fields. As stressed in \cite{belvedere}, this approach introduces a
                redundant Bose field algebra containing some unphysical degrees of
                freedom.
                Therefore some care must be taken to select the fields in the bosonized
                sector needed for the description of the original theory. The redundant
                Bose fields constitute a set of pairwise  massless fields quantized
                with opposite metrics and the appropriate treatment in order to define
                the  correct Hilbert space of states was undertaken in
                \cite{belvedere} in the case of two fermion MT like model with quartic
                interaction only among different species. In the GMT case, under
                consideration here, these features are reproduced according to an
                affine $su(3)$ Lie algebraic constructions.

                We will show that in the bosonization process of the three fermion
                species GMT theory the semi-classical limit of the intermediate effective
                Lagrangian
                turns out to be the $su(3)$ affine Toda model coupled to matter fields.
                This      intermediate effective action has been written in terms of the
                Wess-Zumino-Novikov-Witten (WZNW) action associated to   $su(3)$
                affine Lie     algebra \cite{jhep}. Therefore, in order to gain insight into the  WZNW      origin of the GMT model we undertake the bosonization process using the
                method of the Abelian reduction of the WZNW   theory to treat the various $U(1)$ sectors in a rather direct and      compact    way such that in the semi-classical limit it reproduces the ATM model   studied in Refs. \cite{jmp, jhep}.

                A positive definite Hilbert space of states ${\cal H}$ is identified as
                a quotient space in the Hilbert space hierarchy emerging in the
                bosonization process, following the constructions of \cite{belvedere}. One has
                that each GMT fermion is bosonized in terms of a Mandelstam
                ``soliton'' operator and a spurious exponential field with zero scale dimension, this   spurious field behaves as an identity in the Hilbert space  ${\cal H}$
                and, so, has no physical effects. Afterwards, a set of generalized
                bosonization rules are established mapping the GMT fermion bilinears into
                the corresponding operators composed of the GSG boson fields.

                The study of these models become interesting since the $su(n)$ ATM
                theories  (see \cite{jmp}-\cite{jhep} and \cite{nucl}-\cite{tension})
                constitute excellent laboratories to
                test ideas about confinement \cite{nucl1, tension}, the role of
                solitons
                in quantum field theories \cite{nucl}, duality transformations
                interchanging solitons and particles \cite{jmp, jhep, nucl}, as well as
                the reduction processes of the (two-loop) Wess-Zumino-Novikov-Witten
                (WZNW) theory
                from which the ATM models are derivable \cite{matter, annals}.
                Moreover,
                the ATM type systems may also describe some low dimensional condensed
                matter phenomena, such as self-trapping of electrons into solitons, see
                e.g. \cite{brazovskii}, tunnelling in the integer quantum Hall effect
                \cite{barci}, and, in particular, polyacetylene molecule systems in
                connection with fermion number fractionization \cite{jackiw}.

                Moreover, it has recently been shown \cite{tension} that the $su(2)$
                ATM model describes the low-energy spectrum of QCD$_{2}$ ({\sl one
                flavor} and $N$
                colors in the fundamental and $N=2$ in the adjoint representations,
                respectively). In connection to this point the $su(n)$ ATM theories may
                be
                relevant in the study of the low-energy sector of {\sl multiflavour}
                QCD$_{2}$ with  $N$ colors.

                The work is organized as follows. In the next section we perform the
                functional integral approach, first, to bilinearize the quartic fermion
                interactions and, second, to make the chiral rotations in order to
                decouple the spinors and the auxiliary fields and  write the effective
                action by means of the Abelian reduction of the WZW theory. In section 3 we
                take the semi-classical limit of the effective action and make the
                identification with the ATM model. In section 4 we proceed with the
                bosonization program and use the operatorial formulation to bosonize all the
                ATM like spinors in the intermediate effective Lagrangian and identify
                the SG type fields which must describe the GMT fermions. Furthermore,
                the unphysical degrees of freedom associated to some decoupled free
                fields are identified. The semi-classical limits of the various quantum
                relationships are taken and compared with the classical results of the ATM
                model. In section 5, the positive definite Hilbert space is constructed
                and the fermion-boson mapping is established providing a set of
                generalized bosonization rules. The conclusions and discussions are presented
                in section 6. The relevant results of the classical GMT/GSG equivalence
                in the context of the ATM master Lagrangian formalism are summarized in
                the Appendix.

                \section{Functional integral approach}

                The two-dimensional massive Thirrring model with current-current
                interactions of $N_{f}$ (Dirac) fermion species is defined by the
                Lagrangian
                density\footnote{Our notations and conventions are: $x^{\pm}=x^{0}\pm
                x^{1};\, \pa_{\pm}=\pa_{0}\pm \pa_{1};\, A^{\pm}=A^{0}\pm A^{1}$;\\
                $\eta^{00}=-\eta^{11}=1;\, \epsilon^{01}=-\epsilon^{10}=1;\, \,
                \g^{\mu}\gamma _{5}=\epsilon^{\mu \nu} \g_{\nu}$
                ;\\
                \br
                \nonumber
                \gamma _{0}=\left(
                \begin{array}{cc}
                0 & 1 \\
                1 & 0
                \end{array}
                \right) ,\qquad \gamma^{1}=\left(
                \begin{array}{cc}
                0 & 1 \\
                -1 & 0
                \end{array}
                \right) , \qquad \gamma _{5}=\gamma^{0}\gamma^{1} =\left(
                \begin{array}{cc}
                -1 & 0 \\
                0 & 1
                \end{array} \right),
                \nonumber
                \er
                so the spinors $\psi^{j}$ are of the form
                $\psi^{j} =\left(
                \begin{array}{c}
                \psi^{j} _{(1)} \\
                \psi^{j} _{(2)}
                \end{array}
                \right).$ Define the dual field $\widetilde{\vp}$ by $\pa_{\mu} \vp(x)=
                \epsilon_{\mu \nu} \pa^{\nu}\widetilde{\vp}(x)$.}
                \begin{eqnarray}
                \label{thirring1}
                \frac{1}{k'}{\cal L}_{GMT}[\psi^{j},\overline{\psi}^{j}]=
                \sum_{j=1}^{N_{f}} \{i\bar{\psi}^{j}\gamma^{\mu}\pa_{\mu}\psi^{j}
                        - m^{j}\,\,{\overline{\psi}}^{j}\psi^{j}\}\,
                        \,- \frac{1}{4}
                \sum_{k,\,l=1}^{N_{f}}\[\hat{G}_{kl}\,J_{k}^{\mu} J_{l\, \mu }\],
                \end{eqnarray}
                where the $m^{j}$'s are the mass parameters, the overall coupling $k'$
                has been introduced for later purposes, the currents are defined by
                $J_{j}^{\mu}\,=\,\bar{\psi}^{j}\gamma^{\mu}\psi^{j}$, and the coupling
                constant parameters are represented by a non-degenerate
                $N_{f}\,$x$\,N_{f}$
                symmetric matrix
                \br
                \label{coupl}
                \hat{G}=\hat{g} {\cal G} \hat{g},\,\,\,\,\hat{g}_{ij}\,=\,g_{i}
                \d_{ij},\,\,\,\,{\cal G}_{jk}={\cal G}_{kj}.
                \er

                For example, in the case $N_{f}=3$ the $g_{i}$'s are some positive
                parameters satisfying, along with the ${\cal G}_{jk}$'s,  the relations
                (\ref{matcon}) and (\ref{qcoupl}) at the classical and quantum levels,
                respectively (the semi-classical
                limit of (\ref{qcoupl}) becomes (\ref{couplsemi}) and this
                can be compared to (\ref{matcon})). The ${\cal G}_{ij}$'s sign define the nature of each
                current-current interaction (attractive or repulsive) \cite{rajaraman}. The
                sign of ${\cal G}_{ij}$ is the same as the one for $g_{ij}$ in
                (\ref{matrix0}).

                The GMT model (\ref{thirring1}) is related to the weak coupling sector
                of the $su(n)$ ATM theory in the classical treatment of Refs.
                \cite{jmp, jhep} (see appendix A). We shall consider the special case of
                $su(3)$ ($N_{f}=3$). In the $N_{f}=3$ case the currents at the quantum
                level must satisfy
                \br
                \label{currelat}
                J_{3}^{\mu}\,=\, \hat{\d}_{1} J_{1}^{\mu} + \hat{\d}_{2} J_{2}^{\mu},
                \er
                where the $\hat{\d}_{1,\,2}$ are some parameters related to
                the   couplings $\hat{G}_{kl}$. Notice that the fermion bilinears in  the  constraint (\ref{currelat}) are defined in terms of
                point splitting. Below we will explain that Eq.
                (\ref{currelat}) is
                necessary in order to reproduce the various particle/soliton
                correspondences and will be consistently defined at the level
                of a     quantum field  theory for a field
                sub-algebra. The quantization of constrained non-Abelian fermion    theories with current-current interactions and their relation
                to level $k=2N_{f}$ WZNW model has been considered in the
                literature (see, e.g. \cite{lohe} and references therein). The
                classical counterpart of the currents relationship
                (\ref{currelat}), according to the Lie algebraic construction
                of the $su(3)$ ATM model, is given in (\ref{curclas}).

                Taking into account that the {\sl signs} of the ${\cal
                G}_{ij}$'s in the model (\ref{thirring1}) are equal to the signs of the $g_{ij}$'s in (\ref{matrix0}) ($g_{i}>0$) one can infer that the fermions of the same species will experience an attractive force. The pair of fermions of species  $1$ and $3$, as well as $2$ and $3$ also experience attractive forces,  whereas the pair of fermions $1$ and $2$ suffer a repulsive force
                \cite{rajaraman}. These
                features can also be deduced from the behavior of the time delays due   to soliton-soliton interactions in the associated $su(3)$ ATM model
                studied in Ref. \cite{bueno}.

                In this paper we perform a detailed study of the $N_{f}=3$
                case, however,  the construction below until Eq. (\ref{chiral1}) is valid for any $N_{f}$. In
                the context of the operator formulation the set of fundamental local
                field operators is given by ${\cal F}\equiv {\cal F}\{\bar{\psi}_{j},\,
                \psi_{j}\}$ and the Hilbert space ${\cal H}$ of the theory is constructed
                as a representation of the intrinsic field algebra:  ${\cal H}\dot{=}
                {\cal F}|0>$.  In the functional  integral approach the space ${\cal
                H}$ can be constructed from the generating functional given by
                \begin{eqnarray}
                \label{generating}
                Z_{GMT}[\bar{\theta}_{j},\,\theta_{j}]\,=\, {\cal N}^{-1} \int
                D\bar{\psi} D\psi e^{i W[\bar{\psi}_{i},
                \psi_{i}, \bar{\theta}_{i}, {\theta}_{i}]}
                \end{eqnarray}
                where $W[\bar{\psi}_{i}, \psi_{i}, \bar{\theta}_{i}, {\theta}_{i}]$ is the action in the presence of Grassmannian valued sources
                $\bar{\theta}_{i}$ and ${\theta}_{i}$,
                \begin{eqnarray}
                \label{action}
                W[\bar{\psi}_{i}, \psi_{i}, \bar{\theta}_{i}, {\theta}_{i}]\,=\, \int   d^2 x \[ {\cal L}_{GMT} + \bar{\psi}_{i}{\theta}_{i}+
                \bar{\theta}_{i}\psi_{i}\].
                \end{eqnarray}

                In the next steps we closely follow the procedure adopted in
                \cite{belvedere}. As a first step in the bosonization of the model and    in order
                 to eliminate the quartic interactions, we introduce the ``auxiliary''   vector fields $a_{k}^{\mu}$ in (\ref{generating}) in the form
                \begin{eqnarray}
                \nonumber
                Z^{\prime}_{GMT}[\bar{\theta}_{j},\,\theta_{j},\,\zeta_{j}^{\mu}]&=&
                {\cal N}^{-1} \int D\bar{\psi}\, D\psi\,
                Da_{i}^{\mu}\, \ \mbox{exp} \[i W + i \int d^2 x\{ \sum_{k,\,l} {\cal
                G}_{kl}^{-1} a_{k}.a_{l} + \\
                &&\sum_{k} a_{k}.\zeta_{k}\}\]\label{generating1}
                \end{eqnarray}
                where the ${\cal G}_{kl}^{-1}$'s are the elements of the inverse of the   matrix ${\cal G}$ defined in (\ref{coupl}). In this way  we define an  extended field algebra  ${\cal F}'\equiv {\cal F}'\{\bar{\psi}_{j},\,
                \psi_{j},\, a_{k}^{\mu}\}$  and the source terms for the auxiliary fields
                $a_{k}^{\mu}$ were included in order to keep track of the effects of  the bosonization on building the Hilbert space  ${\cal H}'\dot{=}  {\cal
                F}'\{\bar{\psi}_{j},\, \psi_{j},\,a_{k}^{\mu}\}|0>$. We will show that   the bosonized generating functional $Z^{\prime}_{GMT}$ defines an
                extended positive semi-definite  Hilbert space.

                The bosonization follows by reducing the quartic interaction
                to a quadratic action in the Fermi fields through the ``change of  variables''
                \begin{eqnarray}
                \label{change0}
                a^{\mu}_{k}= A^{\mu}_{k}-\frac{1}{2} \sum_{j\,l} {\cal G}_{kj}
                \hat{g}_{jl} J_{l}^{\mu}
                \end{eqnarray}
                such that
                \begin{eqnarray}
                \nonumber
                 \int da_{i}^{\mu}\,\mbox{exp} \[i \int d^2 x\{ \sum_{k,\,l} {\cal  G}_{kl}^{-1} a_{k}a_{l} -\frac{1}{4} \sum_{k,\,l=1}^{N_{f}}
                \hat{G}_{kl}\,J_{k}^{\mu} J_{l\, \mu } \}\]\\
                = \int dA_{i}^{\mu}\,\mbox{exp} \[i \int d^2 x\{ \sum_{k,\,l} {\cal G}_{kl}^{-1} A_{k}A_{l}- \sum_{k} g_{k} J_{k}^{\mu} A_{k\,\mu}\}\].
                \label{change}\end{eqnarray}

                Then the generating functional (\ref{generating1}) can be written with    the effective Lagrangian density given by
                \begin{eqnarray}
                \label{eff}
                \frac{1}{k'}{\cal L}_{eff}\,=\,
                \sum_{j=1}^{N_{f}}\{i\bar{\psi}^{j}\gamma^{\mu} D_{\mu}(A_{j})\psi^{j}  -   m^{j}\,\,{\overline{\psi}}^{j}\psi^{j}\} + \sum_{j\,k} {\cal
                G}^{-1}_{jk} A_{j}^{\mu} A_{k\, \mu}\,,
                \end{eqnarray}
                where $D_{\mu} (A^{j})\,=\,i \partial_{\mu}- g_{j} A^{j}_{\mu}$\, (no  sum  in $j$).

                Notice that the Lagrangian (\ref{eff}) is local gauge non-invariant due                 to the presence of the terms in the last summation. Since the   $A^{\mu}_{j}$'s are two-component vector fields (in two dimensions) we
                introduce
                the parameterizations $A^{j}_{\pm}$ in terms of the $U(1)$-group-valued   Bose fields ($U_{j}, V_{j} $) as
                \begin{eqnarray}
                \label{change1}
                 A^{j}_{+} \,=\, \frac{2}{g_{j}} U^{-1}_{j}i
                \pa_{+}U_{j};\,\,\,\,A^{j}_{-} \,=\, \frac{2}{g_{j}} V_{j}i
                \pa_{-}V_{j}^{-1},
                \end{eqnarray}
                 such that
                \begin{eqnarray}
                \label{kinetic}
                \bar{\psi}^{j}\gamma^{\mu} D_{\mu}(A_{j})\psi^{j}\,=\,
                (V^{-1}\psi_{j}^{(1)} )^{+} (i\pa_{-})(V^{-1}\psi_{j}^{(1)} )+ (U   \psi_{j}^{(2)} )^{+}
                (i\pa_{+})(U \psi_{j}^{(2)} ).
                \end{eqnarray}

                In order to decouple the Fermi and vector fields we perform the fermion    chiral rotations
                \begin{eqnarray}
                \label{rotation}
                \psi_{j}\,=\, \( \begin{array}{c}
                \psi_{j}^{(1)}\\
                \psi_{j}^{(2)} \end{array}\) =   \( \begin{array}{c}
                V_{j} \chi_{j}^{(1)}\\
                U_{j}^{-1} \chi_{j}^{(2)} \end{array}\)\,=\, \Omega_{j} \chi_{j}\,\,\, (\mbox{no sum in j})   \end{eqnarray}
                with the chiral rotation matrix given by  $\Omega_{j}= \frac{1}{2} (1+ \gamma_{5})U^{-1}_{j} + \frac{1}{2} (1- \gamma_{5})V_{j}$.

                Introduce in the functional integral (\ref{generating1}) the
                identities in the form
                \begin{eqnarray}
                \label{iden}
                1\,=\, \int dU_{j}\, [\mbox{det} D_{+}(U_{j})]\, \delta(\frac{g_{j}}{2} A^{j}_{+}- U^{-1}_{j}i \pa_{+}U_{j})\\
                1\,=\, \int dV_{j}\, [\mbox{det} D_{-}(V_{j})] \,\delta(\frac{g_{j}}{2} A^{j}_{-}- V_{j}i \pa_{-}V^{-1}_{j}),
                \end{eqnarray}
                such that the change of variables from  $A^{j}_{\pm}$ to $ (U_{j},  V_{j})$ is performed by integrating over the fields $A^{j}_{\pm}$.

                Next, performing the chiral rotations (\ref{rotation})  and taking into  account the relevant change in the integration measure we can obtain
                \begin{eqnarray}
                \label{measure}
                \Pi_{j=1}^{N_{f}} d\bar{\psi}_{j}\, d\psi_{j}\,  dA_{\pm}^{j}\,=\,
                \Pi_{j=1}^{N_{f}} d\bar{\chi}_{j}\, d\chi_{j}\, dU_{j}\, dV_{j}\, {\cal   J}   (U, V)
                \end{eqnarray}
                with
                \br
                \label{jacobian}
                {\cal J} (U, V)&=& \mbox{exp} \[ -i \sum_{j} \(\Gamma[U_{j}]+
                \Gamma[V_{j}] +i c_{j} \int d^2 x (A^{\mu}_{j}A^{j}_{\mu} ) \)\]\\
                  &=& \mbox{exp} \[ -i \sum_{j} \( \Gamma[U_{j}]+ \Gamma[V_{j}]
                +\frac{4 c_{j}}{g^{2}_{j}} \int d^2 x  U^{-1}_{j} \pa_{+}U_{j}
                V_{j}  \pa_{-}V_{j}^{-1}\)\]
\nonumber
                \er
                where $\Gamma[g]$ - the Wess-Zumino-Witten (WZW) action
                \cite{witten1}- is given by
                \begin{eqnarray}
                \nonumber
                \Gamma[g]=\frac{1}{8\pi}\int d^2x
                \mbox{Tr}(\partial_{\mu}g\partial^{\mu}g^{-1})+\frac{1}{12\pi}\int d^3y \epsilon^{ijk}
                \mbox{Tr}(g^{-1}\partial_{i}g)(g^{-1}\partial_{j}g)(g^{-1}\partial_{k}g),
 \end{eqnarray}
                and appears in (\ref{jacobian}) with negative level. The last term in    (\ref{jacobian}) takes into account the regularization freedom in the
                computation of the Jacobians for gauge non-invariant theories.

                Using the Polyakov-Wiegman identity \cite{polyakov}
                \begin{eqnarray}
                \Gamma[UV]\,=\,  \Gamma[U] + \Gamma[V] + \frac{1}{4\pi} \int d^2 x   (U^{-1} \pa_{+}U) (V \pa_{-}V^{-1}),
                \end{eqnarray}
                and defining the regularization parameter $a_{j}$ as
                \begin{eqnarray}
                \frac{a_{j}}{2\pi}\,=\, \frac{1}{4\pi}-\frac{4c_{j}}{g^{2}_{j}}
                \end{eqnarray}
                the Jacobian (\ref{jacobian}) can be written as
                \begin{eqnarray}
                \label{jacobian1}
                {\cal J} (U, V)&=& \mbox{exp} \[ \sum_{j} \(-i \Gamma[\Sigma_{j}]
                +\frac{i  a_{j}}{2\pi} \int d^2 x  U^{-1}_{j} \pa_{+}U_{j} V_{j}
                \pa_{-}V_{j}^{-1}\)\],
                \end{eqnarray}
                with $\Sigma_{j}=  U_{j} V_{j} $ being a gauge invariant field.

                In the following we shall consider the general case\footnote{Since the
                fermionic pieces are invariant under local gauge transformations one
                can use the ``gauge invariant'' regularization $a_{j}=0$ in the
                computation of the Jacobians.}( $0 \leq a_{j}< 1$). Therefore, the generating functional (\ref{generating1}) can be written
                in terms of the effective action
                \begin{eqnarray}
                W_{eff}&=& W[U, V]+ \sum_{j=1}^{N_{f}} \int d^2 x\,   \[
                i\bar{\chi}^{j}\gamma^{\mu}\pa_{\mu}\chi^{j}
                        - m^{j}\({\chi}^{* j}_{(1)}\chi_{(2)}^{j} \Sigma_{j}^{-1}
                +{\chi}^{* j}_{(2)}\chi_{(1)}^{j} \Sigma_{j}
                \)\],\label{effaction}
                \end{eqnarray}
where
                \begin{eqnarray}
                \nonumber
                W[U, V]\,=\, \sum_{j=1}^{N_{f}} \( -\Gamma[U_{j}V_{j}] +
                \frac{a_{j}}{2\pi} \int d^2 x   (U^{-1}_{j} \pa_{+}U_{j})( V_{j}
                \pa_{-}V_{j}^{-1})\)
                \\
                 - \sum_{k,\, j=1}^{N_{f}} \int d^2x \frac{{\cal
                G}_{jk}^{-1}}{g_{j}g_{k}} ( U^{-1}_{j} \pa_{+}U_{j})( V_{k}
                \pa_{-}V_{k}^{-1}). \label{w}
                \end{eqnarray}

                Notice that in the Abelian case the WZW functional reduces to the free
                action
                \begin{eqnarray}
                \label{free}
                \Gamma[\Sigma]\,=\, \frac{1}{8\pi} \int d^2 x  \pa_{\mu}\Sigma^{-1}
                \pa^{\mu}\Sigma.
                \end{eqnarray}

                In two-dimensions the vector fields can be written as
                \begin{eqnarray}
                \label{vectors}
                A_{\mu}^{j}\,=\, -\frac{1}{g_{j}} \( \epsilon_{\mu\nu} \pa^{\nu}
                \phi_{j} + \pa_{\mu} \eta_{j}\),
                \end{eqnarray}
                which correspond to the parameterizations
                \begin{eqnarray}
                \label{para}
                U_{j}\,=\, e^{\frac{i}{2}( \phi_{j} +\eta_{j})};\,\,\,V_{j}\,=\,
                e^{\frac{i}{2}( \phi_{j} -\eta_{j})}.
                \end{eqnarray}

                The Eqs. (\ref{effaction})-(\ref{w}) taking into account the relations
                (\ref{free})-(\ref{para}) give rise to the effective Lagrangian
                \begin{eqnarray}
                \nonumber
                \frac{1}{k'}{\cal L}_{eff}&=& \sum_{j=1}^{N_{f}}  \[
                i\bar{\chi}^{j}\gamma^{\mu}\pa_{\mu}\chi^{j}
                        - m^{j}\({\chi}^{* j}_{(1)}\chi_{(2)}^{j} e^{-i \phi_{j}}
                +{\chi}^{* j}_{(2)}\chi_{(1)}^{j} e^{i \phi_{j}} \)\] +\\
                &&\frac{1}{2}\sum_{j,k}^{N_{f}} A_{jk} \pa_{\mu} \phi_{j}\pa^{\mu}
                \phi_{k}+\frac{1}{2}\sum_{j,k}^{N_{f}} F_{jk} \pa_{\mu}
                \eta_{j}\pa^{\mu}
                \eta_{k},\label{lageff}
                \end{eqnarray}
where
                \br
                \label{a11}
                A_{jk}&=&\frac{a_{i}-1}{4\pi}\,
                \d_{jk}-\D_{jk},\,\,\,\,\,\,\,\,\,\D_{jk}\,\equiv\, \frac{{\cal
                G}_{jk}^{-1}}{2 g_{j}g_{k}} \\
                \label{f11}
                F_{jk}&=&-\frac{a_{j}}{4\pi}\, \d_{jk}+\D_{jk} ,\,\,\,\,\,\,\,j,k
                =1,2,3,...N_{f}.
                \er

                Notice that the $\phi_{j}$ scalars will be quantized with negative
                metric for ${\cal G}_{jj}^{-1} \geq 0$.

                One can reproduce the sub-algebra $su(2)$ ATM model associated to each positive root of $su(n)$. So, e.g., set the fields labelled by
                $i=2,3,...,N_{f}$ to zero in (\ref{lageff}). If $\phi_{1}= 2
                \phi,\,\,\chi^{1}=\chi,\,\,\,\eta_{1}=\eta,\,\,\,g_{1}=g$, ${\cal G}_{11}= 2$, ${\cal G}_{11}^{-1}= 1/2$, then taking $a_{1}=0$ one has the Lagrangian ($k'=1$)
\begin{eqnarray}
{\cal L}_{eff}= i\bar{\chi}\gamma^{\mu}\pa_{\mu}\chi
                        - m^{1}\({\chi}_{(1)}\chi^{*}_{(2)} e^{2i\phi}+ \mbox{h.c} \)
                -
                \frac{1}{2}A'_{11} (\pa_{\mu} \phi)^2,
                \label{sl2}
                \end{eqnarray}
                where $A'_{11}=(\frac{1}{\pi}+\frac{1}{g^2})$. The Lagrangian
                (\ref{sl2}) appears in the  path integral approach to the massive  Thirring to sine-Gordon mapping \cite{naon}, and it has
                also been considered in \cite{witten0} as a model possessing a massive fermion state despite a chiral symmetry. Moreover, the model
                (\ref{sl2})  describes the low-energy spectrum, as well as some confinement  mechanism in QCD$_{2}$ ({\sl one flavor} and $N$
                colors in the fundamental and $N=2$ in the adjoint representations,  respectively) \cite{tension}. The relevance of the $su(n)$ ATM like
                theories (\ref{lageff})  in the study of the low-energy sector
                of {\sl multiflavour}   QCD$_{2}$ with  $N$ colors deserves a
                further investigation.

The Lagrangian (\ref{lageff}) exhibits the $\(U(1)\)^{N_{f}} \otimes
                \(U(1)_{5}\)^{N_{f}}$ {\sl vector and chiral symmetries}
                \br
                \nonumber
                \eta_{j} \ra \eta_{j},\,\,\,\,\,
                \phi_{j}\, \ra\, \phi_{j} + 2\L^{j},\,\,\,\,\,
                \chi^{j} \ra  e^{i\a^{j}-i \gamma_5 \L^{j} }\,
                \chi^{j},\,\,\,j=1,2,3,...,N_{f};
                \lab{chiral}
                \er
                where $\a^{j}$ and $\L^{j}$ are real independent parameters.

                Associated to the above symmetries one has the vector and chiral
                currents, respectively
                \br
                \label{chiral1}
                j^{k\,\,\mu}\,=\,
                \bar{\chi}^{k}\gamma^{\mu}\chi^{k},\,\,\,\,\,j_{5}^{k\,\,\mu}\,=\,
                \bar{\chi}^{k}\gamma^{\mu}\gamma_{5}\chi^{k}+ 2 \sum_{l}
                A_{kl} \pa^{\mu} \phi_{l}.
                \er

                \section{Semi-classical limit: $su(3)$ ATM model}

                From this point forward we consider the case $N_{f}=3$. Let us consider
                the semi-classical limit of (\ref{lageff}), $g_{i} \rightarrow
                +\infty\,
                (\D_{jk}\rightarrow 0)$, then
                \begin{eqnarray}
                \nonumber
                \frac{1}{k'}{\cal L}_{\mbox{\sl semicl.}}\,=\, \sum_{j=1}^{3}  \[
                i\bar{\chi}^{j}\gamma^{\mu}\pa_{\mu}\chi^{j}
                        - m^{j}\({\chi}^{* j}_{(1)}\chi_{(2)}^{j} e^{-i \phi_{j}}
                +{\chi}^{* j}_{(2)}\chi_{(1)}^{j} e^{i \phi_{j}} \)+\\
                  \frac{a_{j}-1}{8 \pi} (\pa_{\mu} \phi_{j})^2-\frac{a_{j}}{8 \pi}
                (\pa_{\mu} \eta_{j})^2 \].\label{semiclass}
                \end{eqnarray}

                The model (\ref{semiclass}), disregarding the decoupled $\eta_{j}$
                fields and  under certain conditions imposed on the
                fields and parameters, becomes the $su(3)$ ATM model (\ref{atm3}). In
                fact, rescaling the fields $\chi^{j} \rightarrow
                \frac{1}{\sqrt{\lambda}}\chi^{j}$ the model (\ref{semiclass}) is precisely the so-called      $su(3)$ affine Toda model coupled to matter fields (ATM) \cite{jmp,
                jhep}
                provided that we consider the relationships (\ref{confiel}),
                (\ref{m123}) and
                \br
                \label{imposed2}
                m^{j}\equiv m^{j}_{\chi},\,\,\,\,k'\equiv k
                \lambda,\,\,\,\,\frac{1}{24} \equiv
                \frac{\lambda}{8\pi}(1-a_{i}),\,\,\,\, \l \geq
                \frac{\pi}{3},\,\,\,\,k=\frac{\kappa}{2\pi},\,\, \kappa \in \IZ.
                \er

                The ATM model is known to describe the solitonic sector of its
                conformal version (CATM) \cite{bueno}. The ``symplectic quantization'' method
                has recently been applied to the  $su(3)$ ATM model
                and classically the GMT and the GSG models describe the
                particle/soliton sectors of the theory, respectively \cite{jmp, jhep}. The Lagrangian (\ref{semiclass}) can be written in terms of the
                (two-loop) WZNW model for the scalars (Toda fields) defined in the  maximal  Abelian sub-group of $SU(3)$, the kinetic terms for the spinors which   belong to the higher grading sub-spaces of the $su(3)$ affine Lie
                algebra,   plus some scalar-spinor interaction terms \cite{jhep}. In fact, the Eqs. (\ref{effaction})-(\ref{w}) for $g_{i}\rightarrow \infty$ (take   $a_{i}=0$) reproduce the  Eqs. (8.17) or (8.18) of Ref. \cite{jhep}
                provided  that $\epsilon=-1$ and disregarding an overall minus sign of the   Lagrangian.

                From the point of view of the ATM model defined at the  classical level (\ref{atm3}), the terms $\sum_{jk} \D_{jk}
                \pa_{\mu} \phi_{j}\pa^{\mu} \phi_{k}$ as well as the ones proportional to the  {\sl
                regularization parameters} $a_{j}$ in (\ref{lageff}) have a quantum  mechanical origin.

                Moreover, it has been shown that the classical soliton solutions of the
                system (\ref{semiclass}) satisfy the remarkable equivalence (see
                (\ref{equivalence3}))
                \cite{bueno}
                \br
                \label{equivalence}
                \sum_{k=1}^{3} m^{k}_{\chi} \bar{\chi}^{k}\gamma^{\mu}\chi^{k} \equiv
                \frac{1}{3}\epsilon^{\mu \nu}\partial_{\nu}
                [(2m^{1}_{\chi}+m^{2}_{\chi})
                \phi_{1}+(2m^{2}_{\chi}+m^{1}_{\chi})\phi_{2}],
                \er
                where $j^{\mu}_{k}=\bar{\chi}^{k}\gamma^{\mu}\chi^{k}$ are the $U(1)$
                currents.

                At the classical level there are only two vector (chiral) currents
                since the $\phi$ fields and parameters ($\a$ and $\L$) satisfy the
                conditions (\ref{confiel}) and (\ref{m123}) \cite{bueno, jmp}. The
                remarkable equivalence
                (\ref{equivalence}) has been verified at the classical level and the
                various
                soliton species (up to $2-$soliton) satisfy it \cite{bueno}. In view of
                the property
                (\ref{equivalence}) it has been argued that the model (\ref{semiclass})
                under the restrictions (\ref{confiel}) and (\ref{m123}) presents some
                bag model like confinement mechanism in which the $\chi^{j}$ spinors
                (``quarks'') can live only in
                the regions where $\pa_{x} \phi_{i} \neq 0$; i.e., inside the SG type
                topological solitons (``hadrons'') \cite{bueno}. In this work we give
                an explanation of this effect in the context of the functional and
                operator bosonization techniques.

                \section{Operator approach}

                As the next step in the hybrid bosonization approach we consider the model (\ref{lageff}) (for $N_{f}=3$) and use the Abelian
                bosonization rules to write the $\chi_{j}$ fields in terms of the bosons $\varphi_{j}$
                \br
                \label{bo1}
                \chi^{j}(x) &=& (\frac{\mu}{2\pi})^{1/2} e^{-i\pi\g_{5}/4}
                :e^{i\sqrt{\pi}\(\g_{5}\vp^{j}(x) +
                \int_{x^{1}}^{+\infty}\dot{\vp^{j}}(x^{0},z^{1}) dz^{1} \)}:
                \\
                \label{bo2}
                i\bar{\chi}^{j}\gamma^{\mu}\pa_{\mu}\chi^{j}&=& \frac{1}{2}
                (\pa_{\mu}\vp^{j})^2,\\
                \label{bo3}
                 {\chi}_{(1),\,j}^{*}(x)\chi_{(2)\, j} (x)&=& -\frac{c \mu}{2\pi}
                :e^{i\sqrt{4\pi} \vp^{j}(x)}:\,,\\
                \label{bo4}
                : \bar{\chi }^{j}\gamma ^{\mu }\chi^{j} : &=&-\frac{1}{\sqrt{\pi}}
                \epsilon^{\mu \nu}\partial_{\nu }\vp^{j}
                \er
                where the normal ordering denoted by $:\,:$ is performed with respect to the
                mass $\mu$ which is used as an infrared cut-off and
                $c=\frac{1}{2} \mbox{exp}(\g) \sim 0.891$.

                Next, let us introduce the fields  $\Phi_{j}$ and  $\xi_{j}$ through
                \br
                \label{canonical}
                \vp_{j}&=&-\frac{1}{\D_{j}} [s_{j} \Phi_{j}-\xi_{j}]\,,\,\,\,\,
                \phi_{j}=-\frac{\sqrt{4\pi}}{\D_{j}}(\xi_{j} - r_{j} \Phi_{j}),\\
                \label{canonical1}
                \D_{j}&=&\sqrt{4\pi}(s_{j}-r_{j}),
                \er
                where $s_{j}$ and $r_{j}$ are real parameters. With the fields
                $\Phi_{j}$ defined in (\ref{canonical}) the `mass' terms in
                (\ref{lageff})
                bosonize to the usual `Cos$(\Phi_{j})$' fields in the GSG type models
                \cite{nucl1, jhep}. Then the Lagrangian (\ref{lageff}) in terms of
                purely bosonic fields
                becomes
                \begin{eqnarray}
                \nonumber
                \frac{1}{k'}{\cal L}_{eff}^{'}&=& \sum_{j, k = 1}^{3} \frac{1}{2}
                \[C_{jk}\, \pa_{\mu} \Phi_{j}\pa^{\mu} \Phi_{k} +2 D_{jk}\, \pa_{\mu}
                \xi_{j}\pa^{\mu} \Phi_{k} + E_{jk}\,  \pa_{\mu} \xi_{j}\pa^{\mu}
                \xi_{k}\\
                \label{lageff1}
                &&+ F_{jk}\, \pa_{\mu} \eta_{j}\pa^{\mu} \eta_{k} \]  + \sum_{j=1}^{3}  M^{j} \mbox{cos}(\Phi_{j}),
                \end{eqnarray}
                where
                \br
                \label{c11}
                C_{jk}&=&\frac{1}{\D_{j}^2}[s_{j}^2+(a_{j}-1)r_{j}^2]\, \d_{jk}-4\pi
                \frac{r_{j}r_{k}}{\D_{j}\D_{k}} \D_{jk},\\
                \label{d11}
                D_{jk}&=&-\frac{1}{\D_{j}^2}[s_{j}+(a_{j}-1)r_{j}]\, \d_{jk}+4\pi
                \frac{r_{k}}{\D_{j}\D_{k}} \D_{jk},\\
                \label{e11}
                E_{jk}&=&\frac{a_{j}}{\D_{j}^2}\, \d_{jk}-4\pi
                \frac{\D_{jk}}{\D_{j}\D_{k}},\,\,\,\,\,\,
                M^{j}=\frac{c\, \mu\, m^{j}}{\pi},
                \er
                with the $\D_{jk}$'s defined in (\ref{a11}).

                As the result of the choices (\ref{canonical})-(\ref{canonical1}) it
                emerges an interesting feature. Rescaling the fields $\xi_{j}
                \rightarrow (s_{j}-r_{j}) \xi_{j}'$ in (\ref{lageff1}) one notices that
                the
                symmetric matrices $E_{jk}$, Eq. (\ref{e11}), and $F_{jk}$, Eq.
                (\ref{f11}), are related by an {\sl opposite} sign. Consider the fields  $\xi_{j}''=\sum_{k} U^{jk} \xi_{k}'$ and $\eta_{j}^{'}=\sum_{k}
                U^{jk}\eta_{k}$, where $U$ is an orthogonal matrix which diagonalize  the matrices
                $E$ and $F$ such that the relevant kinetic terms for the fields  $\xi_{j}''$ and  $\eta_{j}^{'}$ are {\sl diagonal}. The new fields
                $\xi_{j}''$
                and  $\eta_{j}^{'}$ will be quantized with opposite metrics. As  considered in \cite{belvedere} the emergence of these decoupled Bose fields
                poses a structural problem related to the fact that the fields $\xi_{j}$    and $\eta_{j}$ do not belong to the field algebra ${\cal F}'$ and
                cannot be defined as operators on the space ${\cal H}'$. Nevertheless, there    are some relevant combinations of them, as we will see below, which   belong to ${\cal H}'$.

                The GMT model for $N_{f}=3$ describes three fermion species with the
                currents constraint (\ref{currelat}) and we are
                faced here with the problem of choosing the corresponding bosonic
                fields that must describe these fermionic degrees of freedom in the
                effective
                bosonic Lagrangian (\ref{lageff1}). On the other hand, in \cite{jmp,
                jhep} by means of the ``symplectic quantization'' method it has been
                shown that
                the three bosonic fields in order to describe the relevant fermions
                (solitons)
                of the three species GMT model must satisfy certain relationship. This
                fact is expressed in the
                restrictions  (\ref{confiel}) and (\ref{m123})  to be imposed on the
                ATM classical model
                (\ref{atm3}) which remains unchanged in the reduced GSG theory
                (\ref{sine3}) \cite{jmp,
                jhep}. This suggests that we must impose an analogous restriction at
                the
                quantum level, thus let us write
                \br
                \label{relation}
                \Phi_{3}= \d_{1} \Phi_{1} + \d_{2} \, \Phi_{2},
                \er
                where the parameters $\d_{1,\,2}$ are determined from the consistency
                conditions imposed for the decoupling of the fields $ \Phi_{j}$ and
                $\xi_{j}$. In fact, once the relationship (\ref{relation}) is assumed
                the terms
                with the $D_{ij}$ coefficients in (\ref{lageff1}) can be written as
                \br
                \nonumber
                \[(D_{11}+\d_{1} D_{13}) \pa_{\mu}\xi_{1}+(D_{21}+\d_{1} D_{23})
                \pa_{\mu}\xi_{2}+(D_{31}+\d_{1} D_{33}) \pa_{\mu}\xi_{3}\]\pa^{\mu}
                \Phi_{1}+&&
                \\
                \nonumber
                \[(D_{12}+\d_{2} D_{13}) \pa_{\mu}\xi_{1}+(D_{22}+\d_{2} D_{23})
                \pa_{\mu}\xi_{2}+(D_{32}+\d_{2} D_{33}) \pa_{\mu}\xi_{3}\]
                \pa^{\mu}\Phi_{2}.&&\\
                \label{decouple}
                \er
                Consider
                \br
                \label{parameteres1}
                \frac{s_{i}}{r_{i}}&=&1-a_{i}+
                4\pi(\D_{ii}-\frac{\D_{ij}\D_{ik}}{\D_{jk}}),\,\,\,i\neq j \neq
                k;\,\,\,\,\,\,i,j,k=1,2,3.\\
                \label{parameteres2}
                \d_{p}&=&-\frac{4\pi \D_{12} \D_{33}-a_{3} \D_{12}-4\pi \D_{31}
                \D_{23}}{4\pi \D_{q3} \D_{pp}-a_{p} \D_{q3}-4\pi \D_{12}
                \D_{p3}},\,\,\,\,p\neq q;\,\,\,\,\,\,p,q=1,2.
                \er

                For the relationships (\ref{parameteres1})-(\ref{parameteres2}) the
                fields $ \Phi_{j}$ and $\xi_{j}$ decouple since all the coefficients in
                (\ref{decouple}) vanish identically.
                Then, with this choice of parameters the Lagrangian (\ref{lageff1})
                becomes
                \begin{eqnarray}
                \nonumber
                \frac{1}{k'}{\cal L}_{eff}^{'}&=& \sum_{j, k = 1}^{3}
                 \frac{1}{2} \[ C_{jk}\, \pa_{\mu}\Phi_{j} \pa^{\mu} \Phi_{k} + E_{jk}\,  \pa_{\mu} \xi_{j}\pa^{\mu} \xi_{k} +
                F_{jk}\, \pa_{\mu} \eta_{j}\pa^{\mu} \eta_{k} \]+\\
\label{lageff2}
                &&\sum_{j=1}^{3}  M^{j}
                \mbox{cos}(\Phi_{j}),
                \end{eqnarray}
                where the parameters $C_{jk}$ can be written as
                \br
                \label{cjj}
                C_{jj}&=& \frac{1}{\b_{j}^2}+ C_{jj}';\,\,\,\,j=1,2,3;\\
                 C_{jj}'&=& -
                \frac{\D_{jl}\D_{jm}}{\D_{lm}}\frac{1}{(\frac{s_{j}}{r_{j}}-1)^2};\,\,\,\,l \neq m \neq j\\
                \label{cjk}
                C_{jk}&=&-\frac{\D_{jk}}{(\frac{s_{j}}{r_{j}}-1)(\frac{s_{k}}{r_{k}}-1)};\,\,\,\,\,\,j\neq  k\\
                \label{betai}
                \b_{j}^2&\equiv& \frac{4\pi - \frac{a_{j}}{{\cal G}^{j}_{lm}}
                g_{j}^2}{1+\frac{g^2_{j}}{\pi} \frac{1-a_{j}}{4{\cal G}^{j}_{lm}
                }};\,\,\,\,\,\,\,l \neq m \neq j,\\
                \label{jlm}
                {\cal G}^{j}_{lm}&\equiv&  {\cal G}^{-1}_{jj}-\frac{{\cal
                G}^{-1}_{jl}{\cal G}^{-1}_{jm} }{{\cal G}^{-1}_{lm}},\,\,\,\,\,\, \frac{s_{k}}{r_{k}}
                = \frac{\frac{\b_{k}^2}{4\pi}}{1-\frac{\b_{k}^2}{4\pi}}+1.
                \er

                It is convenient to make the change
                \br
                \label{rescale}
                \Phi_{j} \rightarrow \b_{j} \Phi_{j}
                \er
                in all the relevant expressions. Therefore, the relationship
                (\ref{relation}) becomes
                \br
                \label{relation1}
                \b_{3}\,\Phi_{3}= \d_{1}\,\b_{1}\, \Phi_{1} + \d_{2} \,\b_{2}\,
                \Phi_{2},
                \er
                where
                \br
                \label{deltasb}
                \d_{1}=-\frac{\D_{12}}{\D_{23}}
                (\frac{\b_{3}^2}{\b_{1}^2})\frac{1-\frac{\b_{1}^2}{4\pi}}{1-\frac{\b_{3}^2}{4\pi}};\,\,\,\,\,\,\d_{2}=-\frac{\D_{12}}{\D_{13}}
                (\frac{\b_{3}^2}{\b_{2}^2}) \frac{1-\frac{\b_{2}^2}{4\pi}}{1-\frac{\b_{3}^2}{4\pi}}.
                \er

                Here we point out a remarkable result. One can verify
                \br
                \label{vanish}
                \frac{1}{2} \sum_{j} C_{jj}' \b_{j}^2 (\pa_{\mu} \Phi_{j})^2 +
                \sum_{j<k} \b_{j}\b_{k} C_{jk} \pa_{\mu}\Phi_{j}\pa^{\mu} \Phi_{k}\,\equiv\,0,
                \er
                in the Lagrangian (\ref{lageff2}); i.e. the coefficient of each
                bilinear term of type $\pa_{\mu} \Phi_{j}\pa^{\mu} \Phi_{k},\,j,k=1,2$ in
                (\ref{vanish}) vanishes identically when the relationship (\ref{relation1})
                and the parameters defined in (\ref{cjj})-(\ref{jlm}) are taken into
                account. This result is achieved for any set of the regularization
                parameters $a_{i}$.

                Then the Lagrangian (\ref{lageff2}) becomes (set $k'=1$)
                \begin{eqnarray}
                \nonumber
                {\cal L}_{GSG}&=& \sum_{j= 1}^{3}
                \[\, \frac{1}{2}\pa_{\mu}\Phi_{j} \pa^{\mu} \Phi_{j} + M^{j}
                \mbox{cos}(\b_{j} \Phi_{j})\] +\\
                &&\frac{1}{2}\sum_{j,\,k =1}^{3} \[E_{jk}\,  \pa_{\mu} \xi_{j}\pa^{\mu}
                \xi_{k}
                + F_{jk}\, \pa_{\mu} \eta_{j}\pa^{\mu} \eta_{k} \],
                \label{lageff3}
                \end{eqnarray}
                with the fields $\Phi_{j}$ satisfying the constraint (\ref{relation1}).
                Thus in (\ref{lageff3}) one has the GSG theory for the fields
                $\Phi_{j}$ and the kinetic terms for the $\xi_{j}$ and $\eta_{j}$ free
                fields, respectively; which completely decouple from the SG fields
                $\Phi_{j}$.

                Notice that the form of the parameter $\beta_{j}$ has been determined
                by requiring the decoupling of the set of fields ($\Phi_{j}$,\,
                $\xi_{j}$)\, and the absence of the ``off-diagonal'' kinetic terms for the
                $\Phi_{j}$ fields in (\ref{lageff3}) which can always be achieved as a
                consequence of (\ref{relation1}). Let us mention that the $\b_{j}$'s  will
                also appear in a natural way in (\ref{fermion1}) related to the
                Mandelstam soliton operators.

                Since the potential $\sum_{j=1}^{3} \[- M^{j}
                \mbox{cos}(\b_{j}\Phi_{j})\]$ defined from (\ref{lageff3}) is invariant
                under $\Phi_{j}\rightarrow \Phi_{j} + \b_{j}^{-1}\, 2\pi\, n_{j}$
                ($n_{j} \in \IZ$) and in addition the $\Phi_{j}$'s satisfy (\ref{relation1})
                we have that the $g_{j}$'s and ${\cal G}_{jk}^{-1}$ for any $a_{i}$
                must satisfy
                \br
                \label{qcoupl}
                \frac{n_{1}}{ {\cal G}_{23}^{-1}}
                \frac{\hat{g}_{1}^2}{g_{1}}+\frac{n_{2}}{ {\cal G}_{13}^{-1}} \frac{\hat{g}_{2}^2}{g_{2}}+ \frac{n_{3}}{
                {\cal G}_{12}^{-1}} \frac{\hat{g}_{3}^2}{g_{3}}&=&0,\,\,\,\,n_{j}\in
                \IZ,\,\,\,\, \hat{g}^2_{j}\equiv
                \frac{1-\frac{\b^2_{j}}{4\pi}}{\b^2_{j}},
\er
where $\b_{j}$ is given in (\ref{betai}). An equivalent expression to
                (\ref{qcoupl}) is
\br
\label{deltas}
n_{1} \d_{1}+n_{2} \d_{2}&=&n_{3},\,\,\,\,n_{j}\in \IZ,
\er
where the $n_{j}$'s are associated  to the topological charges in the GSG
theory.

The fermion mass terms
                bosonize to the corresponding $\mbox{cos} \b_{j} \Phi_{j}$ terms, thus
                being the quantum counterpart of the classical on-shell relations
                (\ref{duality31})-(\ref{duality33}). Notice that (\ref{qcoupl}) becomes the
                quantum version of the relationship (\ref{matcon}). See below more on
                this point.

                The parameters $|\L^{j}|$ in (\ref{sine3}) and their dependences on the
                $g_{j}$'s in Eqs. (\ref{strongweak31})-(\ref{strongweak33}) through
                (\ref{gij}) translate at the quantum level to the $\b_{j}^2$'s  defined in
                (\ref{betai}) for any $a_{j}$.

                Notice that the $a_{j}$ dependence of $\b_{j}$ in (\ref{betai}) is
                similar to the one in the ordinary MT theory, up to the ${\cal G}_{lm}^{j}$
                dependence, see e.g.  \cite{belvedere}. For $a_{j}=0$ (``gauge
                invariant'' regularization) one can define from (\ref{betai})
                \br
                \label{betaj}
                \b_{j}^{2}& \equiv & \frac{4\pi}{1+\frac{g_{j}^2}{\pi}\frac{1}{4 {\cal
                G}_{lm}^{j}}},
                \er
                where ${\cal G}_{lm}^{j}$ is defined in (\ref{jlm}).

                In the semi-classical limit $g_{i} \rightarrow \mbox{Large}$, one has
                from (\ref{betaj})\, $\b_{j}^{2} \rightarrow \frac{16 \pi^2{\cal
                G}_{lm}^{j}}{g_{j}^2}$, then (\ref{deltasb}) provides
                \br
                \label{deltaso}
                \d_{p} = - \frac{g_{p}}{g_{3}}\frac{{\cal G}_{12}}{{\cal
                G}_{q3}},\,\,\,\,q\neq p\,\, (p=1,2).
                \er

                In this limit the relations (\ref{relation1}) and (\ref{qcoupl})
                become, respectively
                \br
                \label{relationsemi}
                \frac{1}{{\cal G}_{12}^{-1}} \, \frac{\Phi_{3}}{g_{3}}+\frac{1}{{\cal
                G}_{23}^{-1}} \, \frac{\Phi_{1}}{g_{1}}+\frac{1}{{\cal G}_{13}^{-1}} \,
                \frac{\Phi_{2}}{g_{2}}&=&0\\
                \label{couplsemi}
                \frac{n_{1}}{ {\cal G}_{23}} g_{1}+\frac{n_{2}}{{\cal G}_{13}} g_{2}+
                \frac{n_{3}}{{\cal G}_{12}} g_{3}&=&0,\,\,\,\,n_{j}\in \IZ.
                \er

                The Eq. (\ref{relationsemi}) reproduces the classical relationship
                (\ref{confiel}) with the fields $\Phi_{j}$ and $\phi_{j}$ conveniently
                identified. On the other hand, (\ref{couplsemi}) may reproduce
                (\ref{matcon}) for certain choices of the  $n_{i}$'s and the ${\cal G}_{ij}$'s.

                In order to describe each SG model related to the corresponding $SU(2)$
                sub-group let us set, e.g.,  $j=1$ and take ${\cal G}^{1}_{23} =1/4$  in
                (\ref{betaj}) then \footnote{The semi-classical limit is achieved by
                setting $a_{i}=0$ first and afterwards  $g_{j}\rightarrow \mbox{Large}$, as
                it is observed in the case of MT/SG. In fact, from (\ref{betai}) (take
                ${\cal G}^{1}_{23} =1/4$) the limiting process in the order indicated
                above provides $\b_{1}^{2}=\frac{4\pi^{2}}{g_{1}^{2}}$ in accordance with
                the semi-classical limit of (\ref{mt/sg}).}
                \br
                \label{mt/sg}
                \b_{1}^{2}= \frac{4\pi}{1+\frac{g_{1}^2}{\pi}},
                \er
                which is the standard SG/MT duality \cite{naon, coleman}.

                The bosonized chiral currents (\ref{chiral1}) become
                \br
                j_{5}^{k\,\mu}\,=\, \sqrt{16\pi}\[\frac{a_{k}}{4\pi
                \D_{k}}\pa^{\mu}\xi_{k}-\sum_{j} \frac{\D_{kj}}{\D_{j}}\pa^{\mu}\xi_{j}
                \].
                \er
                One has that the chiral currents of the model (\ref{lageff}) are
                conserved
                \br
                \label{conschiral}
                \pa_{\mu} j_{5}^{k\,\mu}\,=\,0,\,\,\,\,\,\,k=1,2,3,
                \er
                due to the equations of motion for the $\xi_{j}$ fields
                \br
                \label{eqnsxi}
                \frac{a_{k}}{4\pi \D_{k}}\pa^{2}\xi_{k}-\sum_{j}
                \frac{\D_{kj}}{\D_{j}}\pa^{2}\xi_{j}\,=\,0.
                \er

                In the $su(2)$ case, e.g., set $j_{5}^{k\,\,\mu}=0$ ($k=2,3$)
                ($a_{i}=0$),
                then $\pa^2 \xi_{1}=0$ implies $\pa_{\mu} j_{5}^{1\,\,\mu}=0$. This is
                the
                known result of \cite{witten0} in which the field $\xi^{1}$ is
                associated to the conservation of the chiral current and  the field
                $\Phi_{1}$
                to the  zero chirality sector. Thus, through the SG/MT equivalence one
                has a zero-chirality massive Dirac field $\Psi^{1}$ in the physical
                spectrum, whereas
                the spinor $\chi^{1}$ has a non-zero chirality. In the $su(3)$ case
                this
                picture can directly be translated to the relevant fields and currents
                (see below).

                \section{Hilbert space and fermion-boson mappings}

                In order to conclude with the bosonization program we must identify the
                positive definite Hilbert space and construct the generating functional
                in the GSG sector of the theory. With this purpose in mind, let us
                write the fundamental fields $\(\psi^{j},
                A_{j}^{\mu} \)$  in terms of the bosonic fields $\(\xi_{j}, \Phi_{j},
                \eta_{j}
                \)$, thus the Eq. (\ref{vectors}) becomes
                \br
                \label{newvectors}
                A_{\mu}^{j}\,=\,- \frac{\sqrt{4\pi}\,r_{j}\,\b_{j}}{g_{j} \D_{j}}
                \epsilon_{\mu\nu} \pa^{\nu} \Phi_{j} +\ell^{j}_{\mu}
                \er
                where $\ell^{j}_{\mu}$ are longitudinal currents
                \br
                \label{ell}
                \ell^{j}_{\mu} =-\frac{1}{g_{j}}\(
                -\frac{\sqrt{4\pi}}{\D_{j}}\epsilon_{\mu\nu} \pa^{\nu}
                \xi_{j}+\pa_{\mu}\eta_{j}\)\, \equiv\, \pa_{\mu}
                \ell^{j}.
                \er

                In the next steps we will establish the connections between the fields
                $\psi_{j}$ of the GMT model and the relevant expressions of the GSG
                boson fields $\Phi_{j}$ and $\ell^{j}$. The chiral rotations
                (\ref{rotation}) can be
                written as
                \br
                \label{map}
                \psi_{j}&=&\chi_{j} e^{\frac{1}{2}(i\g_{5} \phi_{j}+ i\eta_{j})}.
                \er

                Taking into account the bosonization rule (\ref{bo1}), the canonical
                transformation (\ref{canonical}), the field re-scaling (\ref{rescale}),
                as well as the parameters defined in (\ref{cjj})-(\ref{jlm}) one can
                write the Fermi fields of the GMT model (\ref{map}) in terms of the
                ``generalized''
                 Mandelstam ``soliton'' fields $\Psi^{j}(x)$
                \br
                \label{map2}
                \psi^{j}(x) = \Psi^{j}(x) \s^{j},\,\,\,\,\, j=1,2,3;
                \er
                where
                \br
                \label{fermion1}
                \Psi^{j}(x)& =& (\frac{\mu}{2\pi})^{1/2}\, K_{j} \,\, e^{-i\pi \g_{5}/4}
                :e^{-i\( \frac{\b_{j}}{2} \g_{5} \Phi^{j}(x) +
                \frac{2\pi}{\b_{j}}\int_{x^{1}}^{+\infty}\dot{\Phi^{j}}(x^{0},z^{1})
                dz^{1}
                \)}:\\
                \label{fermion2}
                \s^{j}&=&
                e^{\frac{i}{2}\(\eta_{j}-\frac{\sqrt{4\pi}}{\D_{j}}\widetilde{\xi}_{j}\)}\\
                \label{fermion3}
                &=& e^{-\frac{i}{2} g_{j} \ell^{j}}.
                \er

                In (\ref{fermion1}) the phase factor\footnote{I thank Prof. M.B. Halpern for
                communication on this point.} $K_{j}=\Pi_{i < j} (-1)^{n_{i}}$ ($i,j$ are
                flavor indices; $n_{i}$ is the number of Fermi fields with index $i$ on which $K_{j}$
                acts) is
                included to make the fields $\Psi^{j}$ anti-commuting for different
                flavors \cite{halpern, frishman}.

                Notice that each $\Psi^{j}$ is written
in terms of a non-local
                expression of  the corresponding bosonic field $\Phi^{j}$ and
                the appearance of the couplings $\b_{j}$ in (\ref{fermion1})
                in  the same form as in the standard sine-Gordon construction of the
                Thirring fermions \cite{coleman}; so,  one can refer the fermions
                $\Psi^{j}(x)$ as  generalized SG Mandelstam soliton operators. In the canonical
                construction of the MT/SG equivalence the arguments of the
                exponentials in the components of (\ref{fermion1}) are identified as the space integrals of
                the quantum fermion currents ${\cal J}_{\pm}^{j}$ expressed in terms
                of the bosonic field $\Phi^{j}$ \cite{sazdovic}. By analogy
                with the Abelian case, various `soliton operators' in terms of
                path ordered exponentials of currents have been presented in
                non-Abelian models \cite{lohe}. In the Abelian case, the
                features above seem to be unique to the
                GMT model considered in this work as compared to the one studied in \cite{belvedere} in which the bosonized fermions do not have the $\b_{j}$
                coupling dependence as in (\ref{fermion1}). In fact, in the
                bosonization of the two species MT like model with quartic interaction only among different species, considered in \cite{belvedere}, the fermion analog  to $\Psi^{j}(x)$ is expressed as a product of two fields with Lorentz
                spin $s=\frac{1}{4}$.

                On the other hand, taking into account $J_{3}^{\mu}\,=\, \hat{\d}_{1}
                J_{1}^{\mu} + \hat{\d}_{2} J_{2}^{\mu}$\, from Eq. (\ref{currelat})
                for
                \br
                \label{hdel}
                \hat{\d}_{p}\,=\,\frac{g_{p}}{g_{3}}\frac{{\cal G}_{12}}{{\cal
                G}_{q3}};\,\,\,\,p\neq q;\,\,\,\,p,q = 1,2
                \er
                one can re-write (\ref{change0}) as
                \br
                \label{vectors1}
                a_{p}^{\mu}&=& A_{p}^{\mu}-\frac{1}{2} \({\cal G}_{pp} g_{p}+{\cal
                G}_{p3} g_{3} \hat{\d}_{p}\) J^{\mu}_{p};\,\,\,\,\,p=1,2\\
                \label{vectors2}
                a_{3}^{\mu}&=& A_{3}^{\mu}-\frac{g_{3}}{2} \({\cal G}_{33}- \frac{{\cal G}_{13}{\cal G}_{23}}{{\cal G}_{12}}\) J^{\mu}_{3}
                \er
                where the currents
                \br
                \label{newcurr}
                J^{\mu}_{k} \equiv {\cal J}^{\mu}_{k} \,=\,
                \bar{\Psi}^{k}\g^{\mu}\Psi^{k};\,\,\,\,k=1,2,3;
                \er
                are written using the relations (\ref{map2}) and
                (\ref{fermion2})-(\ref{fermion3}).

                It is a  known fact that in the hybrid approach to bosonization the
                vectors
                $a_{i}^{\mu}$ are equal to the longitudinal currents \cite{belvedere},
                namely
                \br
                a_{j}^{\mu}= \ell_{j}^{\mu}, \,\,\,\,  j=1,2,3.
                \er

                Then from (\ref{newvectors}) and (\ref{vectors1})-(\ref{vectors2}) one
                can make the identifications
                \br
                \label{currents11}
                {\cal J}^{\mu}_{i}\,=\,-
                \frac{\b_{i}\,\epsilon^{\mu\nu}\pa_{\nu}\Phi_{i}}{2\pi+
                (\frac{a_{i}}{2})\, (g_{i})^2\, [\frac{{\cal G}^{-1}_{jk}}{{\cal
                G}_{jk}}] \mbox{det}({\cal G})};\,\,\,\,\,i\neq j\neq
                k,\,\,\,\,i=1,2,3.
                \er

The form of the current relationship
                (\ref{currents11}) for each related subgroup $SU(2)$ (take $a_{j}=0$) is  exactly the same as the one for the ordinary SG/MT relationship
                \cite{belvedere}. The currents (\ref{newcurr}) written in the form (\ref{currents11})
                when inserted into (\ref{currelat}) reproduce the
                $\epsilon^{\mu\nu}\pa_{\nu}$ derivative of the relationship (\ref{relation1}) between the    boson fields $\Phi_{j}$ for any $a_{j}$. In connection to this
                statement notice that comparing (\ref{parameteres2}) and (\ref{hdel})
                in particular for $a_{i}=0$ one has $\hat{\d}_{p}=-\d_{p}
                \,\,(p=1,2)$. Therefore, Eq. (\ref{currelat}) can be written
                in the form $\pa_{\mu} ({\cal J}^{\mu}_{3}+\d_{1} {\cal
                J}^{\mu}_{1}+\d_{2} {\cal J}^{\mu}_{2})=0$. This expression,
                provided that we assume the relation (\ref{deltas}), is
                the quantum version of (\ref{curclas}) written in the form
                $\pa_{\mu} (J^{\mu}_{3}+\frac{m_{1}}{m_{3}}
                J^{\mu}_{1}+\frac{m_{2}}{m_{3}} J^{\mu}_{2})=0$. Let us emphasize that the classical
                relation (\ref{equivalence3}) holds for the soliton solutions;
                so, each set of choice for $n_{k} \in \IZ $ in the
                corresponding  quantum theory describes the $(n_{1},\,n_{2},\,n_{3} )$ soliton state.

                The ``interpolating'' generating functional (\ref{generating1}) written    in terms of the bosonic fields becomes
                \br
                \nonumber
                {\cal Z}'_{GMT}[\theta^{j}, \bar{\theta}^{j}, \zeta_{k}^{\mu}]&=& {\cal
                N}^{-1} \int D\Phi^{j}\, \d(\b_{3}\Phi_{3}-\d_{1} \b_{1}
                \Phi_{1}-\d_{2} \b_{2} \Phi_{2})\,\, e^{i W[\Phi^{j}]}.\\
                \nonumber
                &&.\int
                D\eta^{j} e^{i W_{0}[\eta^{j}]}
                 \int D\xi^{j} e^{-i W[\xi^{j}]}
                 \mbox{exp}\Big[ i \int d^2 x\\
                && \sum_{k}\{[\bar{\Psi}^{k}
                (\s^{k})^{*}]\theta_{k}+ \bar{\theta}^{k}(\Psi^{k} \s^{k} )
                +\zeta_{k}^{\mu}.\ell^{k}_{\mu}\}\Big],
                \label{generating2}
                \er
                where we have inserted the delta functional to enforce
                (\ref{relation1}). According to (\ref{lageff1}) the actions $W[\eta^{j}]$ and
                $W[\xi^{j}]$ are the free actions for the non-canonical $\eta^{j}$ and
                $\xi^{j}$ fields, respectively, quantized with opposite metrics according to
                the discussion in the paragraph just below Eq. (\ref{e11}) . The action
                $W[\Phi^{j}]$ corresponds to the coupled SG fields  $\Phi^{j}$ in
                (\ref{lageff3}) and the $\Psi^{j}$'s are given in (\ref{fermion1}).

                From  (\ref{generating1}) and (\ref{generating2}) one can get the
                $2n$-point correlation functions for the GMT model (\ref{thirring1}) as
                \br
                \nonumber
                &&<0|\bar{\psi}^{j}(x_{1})...\bar{\psi}^{j}(x_{n})\psi^{j}(y_{1})...
                \psi^{j}(y_{n})|0>'\\
                \nonumber
                &=& <0|\bar{\Psi}^{j}(x_{1})...\bar{\Psi}^{j}(x_{n})\Psi^{j}(y_{1})...
                \Psi^{j}(y_{n})|0>.\\
                &&
                .<0|\s^{*}_{j}(x_{1})...\s^{*}_{j}(x_{n})\s_{j}(x_{1})...\s_{j}(x_{n})|0>_{o},
                \label{correl}
                \er
                where $<0|... |0>$ means average with respect to the GSG theory and
                $<0|... |0>_{o}$ represents average w.r.t. the massless free theories
                $\eta^{j}$ and $\xi^{j}$. The fields $\s_{j}$ give a constant contribution
                to the correlation functions due to the fact that the  $\eta^{j}$ and
                $\xi^{j}$ fields are quantized with opposite metrics, namely
                \br
                <0|\s^{*}_{j}(x_{1})...\s^{*}_{j}(x_{n})\s_{j}(x_{1})...\s_{j}(x_{n})|0>_{o}=1.
                \er

                The auxiliary vector fields $A_{\mu}^{j}$ in
                (\ref{vectors1})-(\ref{vectors2})  belong to the field algebra ${\cal F}'$, and taking into     account that $J_{k}^{\mu} \in {\cal F}'$, one concludes that the
                longitudinal currents $\ell_{j}^{\mu} \in {\cal F}'$.

                The Hilbert space ${\cal H}'$ is positive semi-definite since it has the
                zero norm states
                \br
                <0|\ell^{j}_{\mu}(x) \ell^{j}_{\mu}(y)|0>_{o}=0
                \er
                where the  $(\ell^{j}_{\mu})$'s are the longitudinal currents given in
                (\ref{ell}). These currents generate the field sub-algebra ${\cal
                F}_{o}\equiv {\cal F}_{o}\{\ell_{j}^{\mu}\}$ related to the zero norm states
                ${\cal H}_{o}\dot{=}{\cal F}_{o}|0>\, \subset\,  {\cal H}'$. The
                potential fields $\ell_{j}$ do not belong to ${\cal F}'$, only their
                space-time derivative occur in ${\cal F}'$; in addition, the fields $\s_{j}$
                also do not belong to  ${\cal F}'$. Therefore, the positive semi-definite
                Hilbert space   ${\cal H}'$  is generated from the field algebra
                ${\cal F}'\{\bar{\psi}_{j},\, \psi_{j},\, A_{k}^{\mu}\}={\cal
                F}'\{\bar{\Psi}^{j}\s_{j}^{*} ,\, \Psi^{j}\s_{j},\, \ell_{k}^{\mu}\}$.

                In this way, we make the fermion-boson mapping between the GMT and GSG
                theories in the Hilbert sub-space of states  ${\cal H}'$. For any
                global gauge-invariant functional $F\{\bar{\psi}_{j},\, \psi_{j}\} \in {\cal
                F}$, one can write the one-to-one mapping
                \br
                <0|F\{\bar{\psi}_{j},\, \psi_{j}\} |0>'\equiv <0|F\{\bar{\Psi}^{j},\,
                \Psi^{j}\}|0>.
                \er

                Therefore, one can establish the equivalence
                \br
                Z_{GMT}^{\prime}[\bar{\theta}, \theta, 0] \sim Z_{GMT}[\bar{\theta},
                \theta] \sim Z^{\Phi_{j}}[\bar{\theta}, \theta]
                \er
                with
                \br
                \nonumber
                Z^{\Phi_{j}}[\bar{\theta}, \theta]&=& {\cal N}^{-1} \int  D\Phi^{j}\,
                \d(\b_{3}\Phi_{3}-\d_{1} \b_{1} \Phi_{1}-\d_{2} \b_{2} \Phi_{2})\,\,
                e^{i W[\Phi^{j}]}
                \\
                \label{hilbert}
                &&
                \mbox{exp}\Big[ i \int d^2 x \sum_{k}\{\bar{\Psi}^{k}\theta_{k}+
                \bar{\theta}^{k}\Psi^{k}\}\Big],
                \er
                and the $\Psi^{j}$'s are given in (\ref{fermion1}).  Therefore, the GMT and GSG mapping is established in a positive-definite
                Hilbert space.

                Some comments are in order here. The fields $ \Psi^{j}(x)$ represent
                the physical fermions of the GMT model. In fact, the original spinor
                fields $\psi^{j}$ are bosonized in terms of the  $ \Psi^{j}(x)$ fields
                and
                the exponential operators with zero scale dimension. These spurious
                fields $\s^{j}$ have no physical effects and behave as an identity
                in the Hilbert space of states since the fields $\eta_{j}$ and
                $\xi_{j}$
                are quantized with opposite metrics. On the other hand, according to
                the discussion in the paragraph just below Eq. (\ref{eqnsxi}) and as a
                consequence of the results Eqs. (\ref{conschiral})-(\ref{eqnsxi}) one can
                conclude that the fields
                $\Psi^{i}$ have zero chirality and become massive; whereas, the fields
                with
                non-zero chirality $\chi^{i}$, whose current conservation laws are
                associated to the fields $\xi_{j}$ ($\notin {\cal F}'$), disappear from the spectrum of the theory
                providing a confinement mechanism of their associated degrees of
                freedom. Remember that the fields $\xi_{j}$ and $\eta_{j}$ enter into the spurious fields $\s_{j}$. This picture is the quantum version of the bag model like  confinement mechanism associated to the Noether and topological currents   equivalence (\ref{equivalence}) at the classical level, analyzed in
                \cite{bueno}. This framework also clarifies certain aspects of the  confinement mechanism considered in the $sl(2)$ ATM model at the quantum level  \cite{nucl1, tension}.

                We conclude that it is possible to study the generalized massive   Thirring model (GTM) (\ref{thirring1}) with three fermion species, satisfying   the currents constraint (\ref{currelat}), in terms of the generalized
                sine-Gordon model (GSG) (\ref{lageff3}) with three boson fields,   satisfying  the linear constraint (\ref{relation1}), by means of the   ``generalized'' bosonization rules
                \br
                \label{bosoni1}
                i\bar{\psi}^{j}\g^{\mu}\pa_{\mu}\psi^{j} &=& \frac{1}{2}(1-\rho_{j})
                (\pa_{\mu}\Phi^{j})^2,\,\,\,\,j=1,2,3;\\
                \label{bosoni2}
                 m_{j} \bar{\psi}^{j}\psi^{j}&=& M_{j}\, \mbox{cos}\(\b_{j}
                \Phi^{j}\),\,\,\,\,\,\,\,\,\b_{j}^2=\frac{4\pi}{1+\frac{g_{j}^2}{\pi}\frac{1}{4
                {\cal G}_{lm}^{j}}} \\
                \label{bosoni3}
                \bar{\psi}^{j}\g^{\mu} \psi^{j}&=& -
                \frac{\b_{j}}{2\pi}\,\epsilon^{\mu\nu}\pa_{\nu}\Phi_{j},\\
                \rho_{p}&=& \frac{\b_{p}^2}{2(2\pi)^{2}} \[g_{p}^{2} {\cal
                G}_{pp}-\d_{p}\d_{q}^{-1}\(\sum_{\begin{array}{ccc}
                j&<&k\\
                l&\neq&j\,\neq k \end{array}} g_{j} g_{k} {\cal G}_{jk} \d_{l}
                \epsilon_{l}\)\],\\
                \nonumber
                &&p, q = 1,2; \begin{array}{lll} p&\neq&q,\\
                \d_{3}&=&\epsilon_{3}=\epsilon_{p}=-\epsilon_{q}=1 \end{array}.\\
                \rho_{3}&=& \frac{\b_{3}^2}{2(2\pi)^{2}} \[g_{3}^{2} {\cal
                G}_{33}+\d_{1}^{-1}\d_{2}^{-1}\(\sum_{\begin{array}{ccc}
                j&<&k\\
                l&\neq&j\,\neq k \end{array}} g_{j} g_{k} {\cal G}_{jk} \d_{l}\)\],
                \er
                where the correlation functions on the right hand sides must be
                understood to be computed in the positive definite quotient Hilbert space of
                states ${\cal H} \sim \frac{{\cal H}'}{{\cal H}_{o}}$ defined by the
                generating functional $Z^{\Phi_{j}}[\bar{\theta}, \theta]$ in
                (\ref{hilbert}).

                Let us mention that the WZ term plays a key role in determining the
                fermionic nature of each sine-Gordon type soliton. In fact, by the
                immersion of each $U(1)$  Abelian group into its corresponding $SU(2)$
                sub-group in the bosonized version of the model (\ref{lageff3}) and taking
                into account the relevant WZ term one can proceed as in \cite{eides} to
                determine the fermionic nature of each soliton solution.

                The procedure presented so far can directly be extended to the GMT
                model for  $N_{f}[=\frac{n}{2}(n-1);\,n>3,\,N_{f}$ \,= \,number of positive
                roots of $su(n)$] fermions \cite{nova2}. According to the construction of
                \cite{jhep} (see Appendix) these models describe the weak coupling phase of the
                $su(n)$ ATM models, at the classical level. The strong phase
                corresponds to the GSG theory with $N_{f}[= \frac{n}{2}N_{b},\, N_{b}$ $=(n-1)$ =dimension of the Cartan sub-algebra of $su(n)$] fields, in which
                $\frac{(n-2)(n-1)}{2}$ linear constraints are imposed on the fields.

                \section{Conclusions and discussions}

                Using the mixture of the functional integral and operator formalisms we  have considered the bosonization of the multiflavour
                $N_{f}(=3)$ GMT model with its $U(1)$ currents constrained by (\ref{currelat}). We used the
                auxiliary vector fields in order to bilinearize the various quartic
                fermion interactions. The chiral rotations (\ref{rotation}) decouple
                the spinors from the gauge fields and the Abelian reduction of the WZW
                theory allowed us
                to treat the various $U(1)$ sectors in a rather direct and compact way
                giving rise to the effective Lagrangian (\ref{lageff}). The
                semi-classical limit of the theory at this stage is shown to describe the
                so-called $su(3)$ affine Toda model coupled to matter (ATM)
                (\ref{atm3}); in turn this fact motivated us to impose a relationship
                (\ref{relation}) between the sine-Gordon (SG)
                type fields of the bosonized model  (\ref{lageff1}) in order to
                correctly describe the soliton
                counterparts of the GMT fermions following the results of the classical
                considerations of Refs. \cite{jmp, jhep} (see Appendix). The number of SG type fields
                turns out to be equal to $N_{f}(=\frac{3}{2} N_{b})$. Furthermore, the
                relationship between the SG fields (\ref{relation}) allowed us to
                decouple completely these fields from the remaining bosonic
                fields. The remaining sets of free bosonic fields ($\xi_{j},\,
                \eta_{j}$) are quantized with opposite metrics and their contributions
                are
                essential in order to define the correct Hilbert space of states and
                the relevant fermion-boson mappings. One must emphasize that the
                classical properties of the ATM model motivated the various insights
                considered in the bosonization procedure of the GMT model performed in this
                work. The form of the quantum GSG model (\ref{lageff3}) is similar to its
                classical counterpart (\ref{sine3}), except for the field
                renormalizations and the relevant quantum corrections to the coupling constants.

                Recently, it has been shown that symmetric space sine-Gordon models
                bosonize the massive non-Abelian (free) fermions providing the
                relationships between the fermions and the relevant solitons of the bosonic model \cite{park}. In Abelian bosonization \cite{coleman} there exists an
                identification between the massive fermion operator (charge nonzero
                sector) and a nonperturbative Mandelstam soliton operator; whereas, in
                non-Abelian bosonization \cite{witten1} the fermion bilinears (zero charge
                sectors) are identified with the relevant bosonic operators. In this work
                we have established these type of relationships for interacting massive
                spinors in the spirit of particle/soliton correspondence providing the
                bosonization of the nonzero charge sectors of the GMT fermions by
                constructing the ``generalized'' Mandelstam soliton operators in terms of
                their associated GSG fields, Eq. (\ref{fermion1}). In this way, our work is more close to that of
                \cite{lohe} in which the authors proposed the `soliton operators' as
                exponentials of the non-Abelian currents written in terms of
                bosonic fields, and our constructions may be considered as the relevant Abelian
                reductions. Moreover, in (\ref{bosoni1})-(\ref{bosoni3}) we provide a set of generalized
                bosonization rules mapping the GMT fermion bilinears to relevant bosonic
                expressions which  are established in a positive definite Hilbert space of states ${\cal H}$ .

                On the other hand, the quantum corrections to the soliton masses, the
                bound state energy levels, as well as the time delays under soliton
                scattering in the ATM model, considered in \cite{bueno} at the classical
                level, can be computed in the context of its associated GSG theory
                (\ref{lageff3}).  In addition, the above approach to the GMT/GSG duality may
                be useful to construct the conserved currents and the algebra of the
                corresponding charges in the context of its associated CATM $\rightarrow$
                ATM reduction \cite{nucl}. These currents in the MT/SG case were
                constructed treating each model as a perturbation of a conformal field theory
                (see \cite{kaul} and references therein).

                {\sl Acknowledgments}

                The author is grateful to Prof. L.V. Belvedere for valuable comments and
                Prof. M.B. Halpern for communication about his earlier
                works, Professors V. Juricic and B. Sazdovic for correspondence and enlightening
                comments on their paper, and
                H.L. Carrion, A. Gago, R. Ochoa,
                and M. Rojas  for stimulating discussions on related matters. I
thank Prof. A. Accioly for encouragement and IFT Institute for its
hospitality. The author thanks the hospitality of IMPA and CBPF,
Rio de Janeiro, Brazil, and Prof. M.C. Ara\' ujo at ICET-UFMT.
This work was supported by FAPEMAT-CNPq and in the early stage by
FAPESP.

                \appendix

                \section{$su(3)$ ATM model and GMT/GSG duality}

                 In this Appendix we summarize the Lie algebraic constructions of
                \cite{jmp, jhep} and provide some new results and remarks relevant to our
                discussions.
                The classical aspects of the $su(3)$ ATM model have been considered in
                Refs. \cite{jmp, jhep, bueno}.

                The so-called $su(3)$ ATM Lagrangian is defined by\footnote{In
                \cite{jmp, jhep} the
                ATM model was defined with positive definite kinetic terms for the
                $\phi_{j}$ fields. However, in order to obtain (\ref{atm3}) one can
                consider an overall minus sign in the classical Lagrangian Eq. (2.4) of  Ref. \cite{jmp} taking into account  the reality conditions (2.1)
                in \cite{jmp}. In fact, the $su(2)$ case with single scalar field  $\phi$ has been presented with negative metric \cite{annals, matter,
                tension}.} \cite{jmp, jhep}
                \br
                \label{atm3}
                \frac{1}{k}{\cal L} = \sum_{j=1}^{3} \[ -\frac{1}{24}\(\partial_{\mu
                }\phi_{j}\)^2 + i\overline{\psi}^{j} \gamma ^{\mu}\partial_{\mu
                }\psi^{j} - m_{\psi}^{j} \overline{\psi}^{j} e^{i
                \phi_{j}  \gamma_{5}}\psi^{j}\]
                \er
                where
                $\phi_{1}=\a_{1}.\vp=2\vp_{1}-\vp_{2},\,\phi_{2}=\a_{2}.\vp=2\vp_{2}-\vp_{1},\,\phi_{3}=\a_{3}.\vp=\vp_{1}+\vp_{2}$,\,

                $\a_{3}=\a_{1}+\a_{2}$, $\vp \equiv \sum_{a=1}^{2} \vp_{a} \a_{a}$. The
                $\a_{a}$'s and the $\a_{i}$'s $\,(i=1,2,3)$ are the simple and positive
                roots of $su(3)$, respectively. Consider $\a_{i}^
                2=2$,\,\,$\a_{1}.\a_{2}=-1$. The fields satisfy
                \br
                \label{confiel}
                \phi_{3}=\phi_{1}+\phi_{2}.
                \er

                The soliton type solutions of the model (\ref{atm3}) satisfy the
                remarkable equivalence between the Noether and topological currents
                \br
                \label{equivalence3}
                \sum_{j=1}^{3} m^{j}_{\psi} \bar{\psi}^{j}\gamma^{\mu}\psi^{j} &\equiv
                &
                \epsilon^{\mu \nu}\partial_{\nu}
                (m^{1}_{\psi}\varphi_{1}+m^{2}_{\psi}\varphi_{2}),\\
                \label{m123}
                m^{3}_{\psi}&=&m^{1}_{\psi}+
                m^{2}_{\psi},\,\,\,\,\,\,m^{i}_{\psi}>0.
                \er

                The classical equivalence (\ref{equivalence3}) has recently been
                verified for the
                various soliton species up to $2-$soliton \cite{bueno}.

                The strong/weak couplings dual phases of the model (\ref{atm3}) have
                been uncovered by means of the symplectic and master Lagrangian
                approaches \cite{jmp, jhep}. The strong coupling phase is described by the
                generalized sine-Gordon model (GSG)
                \be
                \label{sine3}
                \frac{1}{k}{\cal L}_{GSG}[\vp]=\sum_{j=1}^{3}\[\frac{1}{24}
                \partial_{\mu}\phi_{j} \partial^{\mu}\phi_{j}
                        +  \;2 m^{j}_{\psi} |\L^{j}| \mbox{cos} \phi_{j}\],
                \ee
                where (\ref{confiel}) must be considered.

                On the other hand, the weak coupling phase is described by the
                generalized massive Thirring model (GMT)
                \begin{equation}
                \label{thirring3}
                \frac{1}{k} {\cal L}_{GMT}[\psi,\overline{\psi}]=
                \sum_{j=1}^{3}\{i\overline{\psi}^{j}\gamma^{\mu}\pa_{\mu}\psi^{j}
                        - m^{j}_{\psi}\,\,{\overline{\psi}}^{j}\psi^{j}\}\,
                        \,- \frac{1}{2} \sum_{k,l=1}^{3} \[g_{kl} J_{k}.J_{l}\],
                \end{equation}
                where $J_{k}^{\mu} \equiv \bar{\psi}^{k}\gamma^{\mu}\psi^{k}$, $g_{kl}$
                are  the coupling constants and the currents satisfy
\br
\label{curclas}
\sum_{j=1}^{3}
                m^{j}_{\psi} \pa_{\mu}\(\bar{\psi}^{j}\gamma^{\mu}\psi^{j}\)
                =0,\,\,\,\,\,\,\,m^{3}_{\psi}=m^{1}_{\psi}+
                m^{2}_{\psi}.
\er

                The signs of the matrix components $g_{ij}$ in (\ref{thirring3})
                according to the construction of \cite{jhep} can be fixed to be
                \br
                \label{matrix0}
                \epsilon_{jk}&\equiv &\mbox{sign}[g_{jk}]
                \\
                \nonumber
                \epsilon_{jk}&=&\Bigg[\begin{array}{cl}
                \mbox{sign}[ (\a_{j})^2], &\,\, j=k\\
                \mbox{sign}[\a_{j}.\a_{k}],&\,\, j \neq k,\,\,\,\,j,k=1,2,3
                \end{array}\Bigg.;\,\,\,\, {\cal \epsilon} = \Bigg(\begin{array}{ccc}
                1& -1& 1\\
                -1& 1& 1\\
                1& 1& 1\\
                \end{array}\Bigg),
                \er
                where the $\a_{i}$'s are the positive roots of $su(3)$.

                It is possible to decouple the $su(3)$ ATM equations of motion obtained
                from the Lagrangian (\ref{atm3}) into the GSG and GMT models equations
                of motion derived from (\ref{sine3}) and (\ref{thirring3}),
                respectively. This is achieved by using the mappings
                \begin{eqnarray}
                \nonumber
                \frac{\psi _{(1)}^{1}\psi_{(2)}^{*\,1}}{i}
                &=&\frac{-1}{4\D}[\left(
                m^{1}_{\psi} p_{1}-m^{3}_{\psi}p_{4}-m^{2}_{\psi}p_{5}\right)
                e^{i\left( \varphi _{2}-2\varphi_{1}\right) }+m^{2}_{\psi}p_{5}e^{3i\left(
                \varphi _{2}-\varphi _{1}\right) }  \\
                &&+m^{3}_{\psi}p_{4}e^{-3i\varphi _{1}}-m^{1}_{\psi}p_{1}]
                \label{duality31}\\
                \nonumber
                \frac{\psi _{(1)}^{2}\psi_{(2)}^{*\,2}}{i}
                &=&\frac{-1}{4\D}[\left(
                m^{2}_{\psi}p_{2}-m^{1}_{\psi}p_{5}-m^{3}_{\psi}p_{6}\right) e^{i\left(
                \varphi _{1}-2\varphi_{2}\right) }+m^{1}_{\psi}p_{5}e^{3i\left( \varphi
                _{1}-\varphi _{2}\right) }+
                \\
                &&
                m^{3}_{\psi}p_{6}e^{-3i\varphi _{2}}-m^{2}_{\psi}p_{2}]
                \label{duality32} \\
                \nonumber
                \frac{\psi_{(1)}^{*\,3}\psi_{(2)}^{3}}{i}
                &=&\frac{-1}{4\D}[\left(
                m^{3}_{\psi}p_{3}-m^{1}_{\psi}p_{4}-m^{2}_{\psi}p_{6}\right) e^{i\left(
                \varphi_{1}+\varphi_{2}\right)}+m^{1}_{\psi}p_{4}e^{3i\varphi _{1}}
                \\
                &&+m^{2}_{\psi}p_{6}e^{3i\varphi _{2}}-m^{3}_{\psi}p_{3}],
                \label{duality33}
                \end{eqnarray}
                where $\D \equiv g_{11}g_{22}g_{33}+2g_{12}g_{23}g_{13}-g_{11}\left(
                g_{23}\right)^{2}-\left( g_{12}\right)^{2}g_{33}-\left( g_{13}\right)
                ^{2}g_{22}$;\,  $p_{1}\equiv \left( g_{23}\right)^{2}-g_{22}g_{33}$;\,
                $p_{2}\equiv \left( g_{13}\right) ^{2}-g_{11}g_{33}$;\, $p_{3}\equiv
                \left(
                g_{12}\right)^{2}-g_{11}g_{22}$;\, $p_{4}\equiv
                g_{12}g_{23}-g_{22}g_{13}$;\, $p_{5}\equiv g_{13}g_{23}-g_{12}g_{33}$;\, $p_{6}\equiv
                -g_{11}g_{23}+g_{12}g_{13}$.

                Moreover, the GSG parameters\,
                $\L^{j}$\, in (\ref{sine3}), the GMT couplings $g_{jk}$ and the
                mass parameters $m^{i}_{\psi}$ in (\ref{thirring3}) are related by
                \begin{eqnarray}
                \L^{1} &=&\frac{-1}{4i\D} \[ m^{3}_{\psi}(g_{12}g_{23}-g_{13}g_{22})+
                m^{1}_{\psi}(g_{22}g_{33}-g_{23}^2)\] , \label{strongweak31} \\
                \L^{2} &=&\frac{-1}{4i\D} \[ m^{3}_{\psi}(g_{12}g_{13}-g_{23}g_{11})+
                m^{2}_{\psi}(g_{11}g_{33}-g_{13}^2)\] , \label{strongweak32} \\
                \L^{3} &=&\frac{-1}{4i\D}
                \[\frac{m^{1}_{\psi}m^{2}_{\psi}}{(m^{3}_{\psi})}
                (g_{13}g_{23}-g_{12}g_{33})+ m^{3}_{\psi}
                ((g_{12})^{2}-g_{11}g_{22})\], \label{strongweak33}\\
                \label{pps}
                m^{3}_{\psi} p_{6} &=& -m^{1}_{\psi} p_{5},\,\,\,\,\,\,\,m^{3}_{\psi}
                p_{4}\, =\, -m^{2}_{\psi} p_{5}.
                \end{eqnarray}

                Following Eq. (\ref{coupl}) let us write
                \br
                \label{gij}
                g_{jk}\equiv \frac{1}{2} g_{j}g_{k} {\cal G}_{jk},
                \er
                then the Eqs. (\ref{m123}) and (\ref{pps}) provide a relationship
                between the matrix elements  ${\cal G}_{jk}$ and the $g_{i}$'s
                \br
                \label{matcon}
                g_{3} {\cal M}_{12}+g_{1} {\cal M}_{23}+g_{2} {\cal M}_{13} =0,
                \er
                where ${\cal M}_{ij}$ is the cofactor of ${\cal G}$.

                Various limiting cases of the relationships
                (\ref{duality31})-(\ref{duality33}) and
                (\ref{strongweak31})-(\ref{strongweak33}) can be taken \cite{jmp}. These relationships incorporate
                each $su(2)$ ATM
                sub-model (particle/soliton)
                weak/strong coupling correspondences; i.e., the ordinary massive Thirring/sine-Gordon relationship \cite{annals}.

                Moreover, the $su(n)$ ATM theory is described by the scalar fields
                $\vp_{a}\, (a=1,...n-1)$ and the Dirac spinors $\psi^{j}$, ($j=1,...N_{f}$;
                $N_{f}\equiv \frac{n}{2}(n-1)$ = number of positive roots\, $\a_{j}$\,
                of the simple Lie algebra $su(n)$) related to the GSG and GMT models,
                respectively \cite{jhep}. From the point of view of its solutions, the
                one-(anti)soliton solution associated to the field\,
                $\phi_{j}=\a_{j}.\vp$\, ($\vp=\sum_{a=1}^{n-1}\vp_{a} \a_{a}$,\, $\a_{a}$= simple roots of     $su(n)$) corresponds to each Dirac field $\psi^{j}$ \cite{jmp, bueno,
                matter}.

\end{document}